\documentclass[12pt,a4paper]{article}
\usepackage{graphicx}
\usepackage{tikz}
\usepackage{a4wide}
\usepackage{comment}

\usepackage{ucs}                % Unicode support in LaTeX
\usepackage[utf8x]{inputenc}    % We use UTF-8 encoded files only (omits all problems with special characters)

\usepackage[T1]{fontenc}     % for the right font encoding
\usepackage{lmodern}         % fonts for good pdf files

\usepackage[british]{babel}
\usepackage{amsmath,amssymb,amsfonts,color}
\usepackage{bm}

\usepackage[sfdefault=cmbr,OMLmathsans]{isomath}  % notation acc. to ISO 31
\usepackage{subfig}
\usepackage{hyperref}
\usepackage{pgfplots}

\usepackage{mathrsfs}

\usepackage{amsthm}
\usepackage{bbm}
\usepackage{upgreek}  % upright greek letters
\usepackage{tensor}  % for tensor index notation

\newcommand{\Gdiff}{\updelta} % Gateaux differential
\newcommand{\mv}{{\pi}} % measure-valued solution (probability measure)
\newcommand{\Vq}{{\V{q}}}
\newcommand{\Vu}{{\V{u}}}
\newcommand{\Vv}{{\V{v}}}

\newcommand{\Eo}{\mathbb{E}} % expectation operator 
\newcommand{\E}[1]{\Eo\bigl[#1\bigr]} % expectation

\newcommand{\Dom}{D} % computational domain
\newcommand{\sU}{\mathcal{U}} % space displacement
\newcommand{\sP}{\mathcal{P}} % space damage/phase-field variable
\newcommand{\rsU}{\mathscr{U}} % stochastic space displacement
\newcommand{\rsP}{\mathscr{P}} % stochastic space damage/phase-field variable

\newcommand{\En}{E} % Energy
\newcommand{\Ens}{\tns{E}} % (stochastic) Energy
\newcommand{\rv}[1]{{\tns {#1}}} % random variable
\newcommand{\rV}[1]{{\tnb {#1}}} % random vector
\newcommand{\uu}{\tnb{u}}
\newcommand{\vv}{\tnb{v}}

\definecolor{darkcyan}{rgb}{0.0, 0.55, 0.55}

\definecolor{light-blue}{rgb}{0.1,0.85,1} 
\definecolor{hgmgrey}{gray}{0.75}

\newcommand{\ignore}[1]{}

\newcommand{\Lp}{\mathrm{L}}

\newcommand{\Hp}{\mathrm{H}}
\newcommand{\Ck}{\mathrm{C}}

\newcommand{\ip}[2]{\left\langle #1 , #2 \right\rangle}
\newcommand{\bkt}[2]{\left\langle #1 \mid #2 \right\rangle}

\newcommand{\ipd}[2]{\left\langle\negthinspace\ip{#1}{#2}\negthinspace\right\rangle}
\newcommand{\bkd}[2]{\left\langle\negthinspace\bkt{#1}{#2}\negthinspace\right\rangle}
\newcommand{\ns}[1]{\left| #1 \right|}
\newcommand{\nd}[1]{\left\Vert #1 \right\Vert}
\newcommand{\nt}[1]{\ns{\negthinspace\ns{\negthinspace\ns{#1}\negthinspace}\negthinspace}}
\newcommand{\trpos}{{\ops{T}}}

\newcommand{\dd}{\mathop{}\!\partial}
\newcommand{\di}{\mathop{}\!\mathrm{d}}

\newcommand{\bbbone}{{\mathchoice {\rm 1\mskip-4mu l} {\rm 1\mskip-4mu l}
{\rm 1\mskip-4.5mu l} {\rm 1\mskip-5mu l}}}

\newcommand{\vek}[1]{\mathchoice{\displaystyle\boldsymbol{#1}}
 {\textstyle\boldsymbol{#1}}{\scriptstyle\boldsymbol{#1}}
 {\scriptscriptstyle\boldsymbol{#1}}}

\newcommand{\ops}[1]{\mathchoice{\displaystyle\mathsf{#1}}
 {\textstyle\mathsf{#1}}{\scriptstyle\mathsf{#1}}
 {\scriptscriptstyle\mathsf{#1}}}

\newcommand{\tnb}[1]{\mathchoice{\displaystyle\mathboldsans{#1}}
 {\textstyle\mathboldsans{#1}}{\scriptstyle\mathboldsans{#1}}
 {\scriptscriptstyle\mathboldsans{#1}}}

\newcommand{\tns}[1]{\mathchoice{\displaystyle\mathsans{#1}}
 {\textstyle\mathsans{#1}}{\scriptstyle\mathsans{#1}}
 {\scriptscriptstyle\mathsans{#1}}}

\newcommand{\skipifemptyarg}[1]{\ifthenelse{\isempty{#1}}{}{\left[#1\right]}}
\newcommand{\skipifscalar}[1]{\ifthenelse{\isempty{#1}}{}{;#1}}

\newcommand{\sL}{\mathrm{L}} % Lp-spaces

\newcommand{\cov}{\mathrm{cov}} % covariance
\providecommand{\C}{\mathcal}% math caligraphic --- caps only
\providecommand{\F}{\mathfrak}% math Fraktur
\providecommand{\MR}{\mathrm}% math Roman
\providecommand{\B}{\mathbb}% math blackboard bold --- caps only
\newcommand{\ch}[1]{\bbbone_{#1}} % characteristic function
\newcommand{\dual}[2]{\langle #1, #2 \rangle}

\renewcommand{\P}{\mathbb{P}} % probability measure
\newcommand{\sQ}{\mathcal{Q}} % parameter space

\newcommand{\sY}{D} % computational domain

\newcommand{\bs}[1]{\boldsymbol{#1}}

\newcommand{\set}[1]{\mathbb{#1}} 
\newcommand{\V}[1]{\bs{#1}} % vector
\newcommand{\x}{{\V{x}}}
\newcommand{\y}{{\V{y}}}

\newcommand{\vep}{\ensuremath{\varepsilon}}
\newcommand{\Vep}{\bm\vep}

\newcommand{\sR}{\set{R}}

\DeclareMathOperator*{\argmin}{arg\,min}
\DeclareMathOperator*{\argmax}{arg\,max}

\newcommand{\D}[1]{\di #1}
\newcommand{\Dm}[2]{\,{#1}(\di #2)}
\newcommand{\Hm}{\mathbb{H}} %Hausdorff measure

\theoremstyle{plain}

\usepackage{authblk}
\date{\today}
\author[1]{Tymofiy Gerasimov}
\author[2]{Ulrich R\"omer}
\author[3]{Jaroslav Vond\v{r}ejc}
\author[3]{\authorcr Hermann G. Matthies}
\author[4]{Laura De Lorenzis}
\affil[1]{Institut f\"ur Angewandte Mechanik,\authorcr Technische Universit\"at Braunschweig, Germany}
\affil[2]{Institut f\"ur Dynamik und Schwingungen,\authorcr Technische Universit\"at Braunschweig, Germany}
\affil[3]{Institute of Scientific Computing,\authorcr Technische Universit\"at Braunschweig, Germany}
\affil[4]{Department of Mechanical and Process Engineering,\authorcr ETH Z\"urich, Switzerland}

\begin{document}

\title{Stochastic phase-field modeling of brittle fracture: computing multiple crack patterns and their probabilities}

\maketitle

\begin{abstract}
In variational phase-field modeling of brittle fracture, the functional to be minimized is not convex, 
{so that} the necessary stationarity conditions of the functional may admit multiple solutions. The solution obtained in 
{an actual computation} is typically one out of several local minimizers.  Evidence of multiple solutions induced by small perturbations of numerical or physical parameters was occasionally recorded but not explicitly investigated in the literature. In this work, we focus on this issue and advocate a paradigm shift, away from the search for one particular solution {towards} the simultaneous description of all possible solutions (local minimizers), along with the probabilities of their occurrence. Inspired by recent approaches 
{advocating} measure-valued solutions 
{(Young measures as well as their generalization to statistical solutions)} and their numerical approximations in fluid mechanics, we 
{propose} the stochastic relaxation of the variational brittle fracture problem {through random perturbations of the functional.} 
We introduce the concept of \emph{stochastic solution}, with the main advantage that point-to-point correlations of the crack phase fields in the underlying domain can be captured. These 
{stochastic solutions are represented by random fields or random variables with values in the classical deterministic solution spaces}. 
{In the numerical experiments, we use a simple Monte Carlo approach to compute approximations to such stochastic solutions.} 
The final result of the computation is not a {single} crack pattern, but rather several possible crack patterns and their probabilities.  The stochastic solution framework using evolving random fields allows additionally the interesting
possibility of conditioning the probabilities of further crack paths on intermediate crack patterns.
\end{abstract}

\vspace{1em}
\vspace{1em}

%\clearpage
\tableofcontents
\vspace{2em}

\section{Introduction}
\label{Intro}

The phase-field approach to brittle fracture dates back to the seminal work of Francfort and Marigo \cite{Francfort1998} on the variational formulation of quasi-static brittle fracture and to the related regularized variational formulation of Bourdin et al.\ \cite{Bourdin2000,Bourdin2007a,Bourdin2007b,Bourdin2008}. The former is the mathematical theory of quasi-static brittle fracture mechanics, which recasts Griffith's energy-based principle as the minimisation problem of an energy functional. The latter presents an approximation, in the sense of $\Upgamma$-convergence, of this energy functional and enables an efficient numerical treatment.

The phase-field formulation of fracture holds a number of advantages over the classical techniques based on a discrete fracture description, whose numerical implementation requires explicit (in the classical finite element method) or implicit (within e.g. the extended finite element method) handling of the discontinuities. The most obvious one is the ability to track automatically a cracking process with arbitrarily complex crack topology, featuring e.g.\ coalescence and branching, also in three dimensions, by describing the evolution of a smooth \emph{crack phase field (which can be interpreted as a damage field)} on a fixed mesh. Another advantage is the ability to describe crack nucleation, also in the absence of singularities, without the need for ad-hoc criteria. Also, by adopting a formulation capable to distinguish between fracture behavior in tension and compression, no supplementary contact problem has to be posed for preventing crack faces interpenetration. For these reasons, phase-field formulations of brittle fracture have attracted a lot of attention in the past decade, see e.g.\ \cite{DelPiero2007,Lancioni2009,Amor2009,Freddi2010,Kuhn2010,Miehe2010a,Miehe2010b,Pham2011,Borden2014,Vignollet2014,Mesgarnejad2015,Kuhn2015,Ambati2015review,Marigo2016,Tanne2018,Sargado2018,Gerasimov2018,WuNguyen2018}. Although several extensions to more complex material behavior and coupled formulations with additional fields (e.g. temperature, concentration etc.) have been proposed, in this work we focus our attention on quasi-static brittle fracture.

\begin{figure}[h]
\begin{center}
\includegraphics[width=1.0\textwidth]{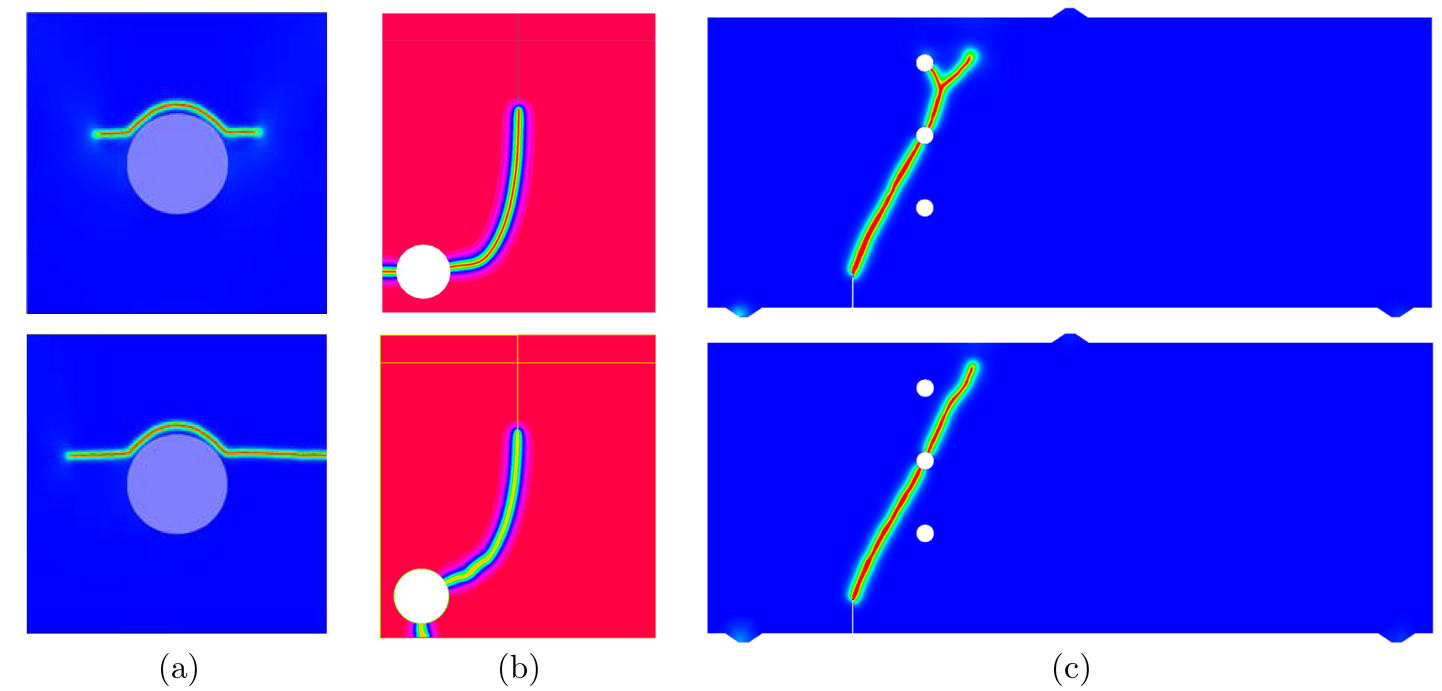}
\end{center}
\caption{Examples of non-unique solutions (in terms of crack path): the fiber-reinforced matrix traction test \cite{Bourdin2008} (similar findings in \cite{Bourdin2000,Bourdin2007a,Amor2009}), (b) the anti-plane shear test \cite{Artina2015} (also in \cite{Burke2010a,Artina2014}), (c) the notched three-point bending test from \cite{Ingraffea1990}.}
\label{fig:NU}
\end{figure}

As typical for solid mechanics problems in presence of softening behavior \cite{Kruzik2019,Mielke2015book}, the functional to be minimized is non-convex with respect to its two arguments (displacement field and crack phase field) simultaneously. This implies that the governing equations of the coupled problem, which are obtained as the necessary conditions of stationarity of the functional with respect to the two arguments, may admit multiple solutions. Since no general numerical algorithm exists which can guarantee global minimization for a non-convex problem, the solution obtained in the computational practice is typically a local minimizer\footnote{Note that the solutions may even correspond to local maxima or saddle points, which can only be clarified by a stability analysis. The questions as to what critical point we actually compute \cite{Bourdin2007a} and to which type of minimizer (global or local) represents a physically meaningful solution \cite{Bourdin2007b} remain beyond the scope of this paper.}. The occurrence of multiple solutions has been occasionally reported in the literature, e.g. in \cite{Bourdin2000,Bourdin2007a,Bourdin2008,Amor2009,Burke2010a,Artina2014,Artina2015}. Some of the examples are illustrated in Figure \ref{fig:NU}. 

Figure \ref{fig:NU}(a) depicts the results obtained in \cite{Bourdin2008} for the traction test on a fiber-reinforced matrix in plane stress. A square elastic matrix is bonded to a rigid circular fiber. The fiber is fixed, while a uniform vertical displacement is applied to the upper edge of the matrix, and the remaining sides are traction free. The computations are performed on an unstructured triangular uniform finite element mesh using the alternate minimization solution algorithm. In the beginning stage of the monotonically increasing loading, a crack nucleates and propagates symmetrically with respect to the vertical axis. However, subsequently the solution looses symmetry and the authors comment that an asymmetric solution ``is consistent with the lack of uniqueness of the solution for the variational formulation''.

Figure \ref{fig:NU}(b) reports two solutions of the anti-plane shear experiment considered in \cite{Artina2015}. The setup, which we also adopt in our numerical experiments in this paper, is detailed in section~\ref{CaseStudy}. In \cite{Artina2015}, two solution algorithms are proposed, which differ in the sequence of alternate minimization and mesh adaptivity: Algorithm 1 ({\em solve-then-adapt}) applies mesh adaptation after the convergence of the minimization procedure, whereas with Algorithm 2 ({\em solve-and-adapt}) mesh adaptivity is carried out at each minimization iteration. The results leading to the top and bottom plots in Figure \ref{fig:NU}(b) are obtained using Algorithm 2 with anisotropic and isotropic adaptive triangulation, respectively.

Finally, Figure \ref{fig:NU}(c) presents our findings for the notched three-point bending experiment originally designed in \cite{Ingraffea1990} with the aim of studying the effect of structural imperfections such as holes on crack trajectories. Our computations performed for one of the setups in \cite{Ingraffea1990} lead to two different solutions. In both cases, the same triangular finite element mesh is used, which is pre-adapted in the region where crack propagation is expected, but we use different increments of the applied displacement. The results shown correspond to the last loading step, featuring in both cases the same magnitude of applied displacement.

The above results suggest the following {\em numerical} factors as possible triggers of multiple solutions:
\begin{itemize}
	\item finite element mesh, including e.g. the choice of fixed vs.\ adaptive, isotropic vs.\ anisotropic mesh,
	\item solution algorithm, e.g.\ hierarchy of multiple nested iterative solution processes or choice of the tangent stiffness matrix in monolithic solution schemes \cite{Gerasimov2016,Wick2017a},
	\item parameters related to the solution algorithm (loading increments, thresholds, tolerances, termination criteria,  etc.),
	\item round-off errors.
\end{itemize}
On the other hand, small perturbations of {\em physical} parameters (geometry parameters or material properties, i.e.\ elastic moduli or fracture toughness) may also lead to multiple solutions. As opposed to the perturbation of numerical parameters, the latter can be interpreted as representing physically meaningful variations in the geometry and material properties which would also be encountered in reality (e.g.\ in experiments). 

Despite the above findings, to the best of the authors' knowledge, no attempt has been made so far to intentionally investigate and characterize the encountered multiple solutions.

In the present work, we aim at addressing the issue of solution non-uniqueness for phase-field modeling of brittle fracture. In doing so we advocate a paradigm shift, from the search for one solution to the simultaneous description of {\em all} possible solutions (local minimizers) along with the probabilities of their occurrence. 

To this end, we shift from a deterministic to a stochastic formulation, so that the multiple deterministic solutions appear as possibilities of a probabilistic solution,
where the different probabilities reflect the energy landscape which may favour
one possibility over another one. This stochastic formulation may be viewed as a relaxation of the original variational problem. Note that the concept of relaxation is not uncommon for non-convex problems. As follows, we outline some of the available approaches.

One relaxation approach builds on parameterized measures 
(Young measures and their generalizations), which are able to describe oscillatory or
concentration effects of minimizing sequences \cite{Roubicek1997book,Pedregal1997,DiPerna1987} in
minimization problems. They are mainly used as a tool for relaxation (generalization) of mathematical formulations which lack minimizers \cite{Benesova2017}. As examples one can name damage evolution in elastic materials \cite{FiaschiEtal12},
micromagnetics and shape memory alloys \cite{Kruzik2001,Kruzik2005}, optimal control \cite{Henrion2019,Kruzik1999}, or fluids \cite{DiPerna1987}. 
A recent inspiring approach focuses on measure-valued solutions and their numerical 
approximations for systems of hyperbolic conservation laws 
\cite{fjordholm2016computation,lanthaler2015computation,Fjordholm2016a,fjordholm2017construction,Fjordholm2018} 
in the context of fluid mechanics. In these papers, the field problem is reformulated with 
Young measure-valued solutions and extended to what is called a ``statistical solution''
--- an infinite system of Young correlation measures ---
which defines a probability measure on a function space, building on a solution concept
which goes back at least to Foia\c{s}, cf.\ \cite{FoiasTemam04} for a concise account.  
This is in fact a special case of a more general construct which has been variously termed a 
``weak distribution'' or ``operational process'' 
\cite{segal56-TAMS,segal58-TAMS,segalNonlin1969}, resp.\ a ``generalised stochastic process''
\cite{gelfand64-vol4} --- a construction which seems to be more suited to infinite dimensional
Hilbert and Banach spaces than the stricter concept of a probability measure,
and one which is inherently connected with a weak or variational problem formulation. This extension or relaxation of the solution space is used in 
\cite{Fjordholm2016a,fjordholm2016computation} to show that in this extended sense
there is a unique solution to certain systems of hyperbolic conservation laws.  
In \cite{fjordholm2016computation,lanthaler2015computation} a Monte Carlo method is used for these systems of hyperbolic conservation laws to sample the solution 
as a random variable in order to compute its mean and variance field,
which then are unique quantities. Due to oscillations of the numerical solution 
on ever finer grids,
such a statistical solution model does not require an underlying 
uncertainty of physical model parameters, although they can be integrated seamlessly.
In the case of brittle fracture which is of interest here, such oscillations do not occur as meshes are refined, it is rather the possible alternative crack
paths we want to capture, and some stochastic perturbation will be introduced.

Beside the inspiration by this idea of statistical solutions, the work we report in this paper also uses methods from
similar work for stochastic problems in mechanics and applied sciences, where a formulation based 
on spectral approximations in a Galerkin setting was proposed in \cite{ghanem1991stochastic},
and then extended in a variational framework to linear elasticity
\cite{hgmCbu99,babuska2004stochastic}, covering also theoretical aspects.  
This was further extended to a variational theory also
for nonlinear problems \cite{matthies2005stochastic}, as well as to thermodynamically
irreversible and highly non-smooth problems of infinitesimal plasticity \cite{HgmRos08a,bvrHgm11}
--- which is in many ways close to the present case of quasi-brittle fracture ---
within the framework of convex analysis \cite{BVRhgm2015}.  In plasticity, uniqueness of the solution
can only be established in the presence of hardening, and is lost
when one considers perfect plasticity.

In this paper we pursue a relaxation 
of the variational problem of phase-field regularized brittle fracture by allowing the solution to be
a random variable with values in the deterministic solution space,
i.e.\ a \emph{stochastic solution}.  
The randomness is introduced through some small random perturbation in the 
energy functional, which becomes itself a random variable.  
The idea is to minimise the \emph{expected value} of the energy functional over
an appropriate space of random fields, and then look at the limiting behaviour of the
stochastic solution as the random perturbation becomes smaller and smaller.

Thus we try to capture all non-unique solutions in a stochastic solution.
Compared to a formulation based on Young measures, such an approach results in solutions which are random fields, and allows to capture point-to-point correlations of the crack phase fields in the underlying domain. This seems also to be at least as general as the idea of a
structure of Young correlation measures  in \cite{Fjordholm2016a,fjordholm2016computation}.
The difference is that we work with random fields or abstract random variables with
values in the deterministic space of possible solution fields, rather than probability measures.
Just as in e.g.\ \cite{babuska2004stochastic,matthies2005stochastic,BVRhgm2015}, this
has certain advantages, as these abstract random variables or random fields live freely
in vector spaces, whereas probability measures are constrained by the requirements
of being positive and necessarily integrating to one, i.e.\ they lie on the intersection
of the positive quadrant (or cone) with the unit ball in some appropriate measure space.
Although the resulting sets of such probability measures are convex, the mentioned constraints usually require particular care when discretisations and numerical approximations
have to be used in the actual computation of an approximate solution.  Formulations
in terms of abstract random variables or random fields on the other hand look completely analogous to the
usual deterministic variational formulations, and thus allow for the use of the
deterministic solver codes; no new equations have to be formulated and new solver
programs to be developed and implemented for probabilistic descriptors like higher moments
\cite{FoiasTemam04}, or systems of Young correlation measures
\cite{Fjordholm2016a,fjordholm2016computation}  for statistical solutions. An important question was whether by this kind of relaxation it would be possible to achieve a unique albeit probabilistic solution, as in
\cite{babuska2004stochastic,matthies2005stochastic,BVRhgm2015}.  As shown
in Section~\ref{secProb_1D}, it seems that uniqueness
is by no means automatic through such a relaxation, and there seems to be
a dependence of the stochastic solution on the kind of perturbation.  Thus it becomes natural to expect that random perturbations be connected with the
ones actually occurring in the physical systems being modelled.
For a given type of perturbation, with the proposed approach the final result of the computation is not a crack pattern (along with the corresponding global and local results, e.g.\ load-displacement curve, displacement field, etc.), but rather several possible crack patterns (with the additional associated results) as well as their probabilities. Such results have obviously a much higher computational cost, however, they also have a much higher information content.

The paper is organised as follows. In Section~\ref{SecPF} we outline the main concepts of phase-field modeling of brittle fracture and the formulation used in the present paper. We adopt the numerical setup in Figure~\ref{fig:NU}(b) taken from \cite{Artina2015}, and in addition to the two crack patterns obtained in the original reference we find a third one by computing on fixed meshes. We term these results deterministic and use them as a motivation and a starting point for the following developments. The stochastic analysis is introduced in Section~\ref{secProb_1D} in a simple one-dimensional setting,  first based on Griffith's theory and then formulated for phase-field modeling. In Section~\ref{secProb_gen} we present the proposed stochastic approach in a rather general setting and apply it to the anti-plane shear case study of Section~\ref{SecPF}. Finally, conclusions are drawn in Section~\ref{Conclusions}.

%%%%%%%%%%%%%%%%%%%%%%%%%%%%%%%%%%%%%%%%%%%%%%%%%%%%%%%%%%%%%%%%%%%%%%%%%%%%%%%%%%%%
\section{Deterministic modeling}
\label{SecPF}

In this section, we briefly recall the main concepts of the phase-field framework for modeling brittle fracture, and then introduce the anti-plane shear test as reference example. For this test we capture three different fracture mechanisms by perturbing the finite element mesh and the loading increment. We also present energy-displacement curves to enable preliminary considerations on the energetic equivalence of the multiple solutions obtained.

%%%%%%%%%%%%%%%%%%%%%
\subsection{Phase-field formulation of brittle fracture}  \label{SSecBrittle}
\begin{figure}[!ht]
\begin{center}
\includegraphics[width=1.0\textwidth]{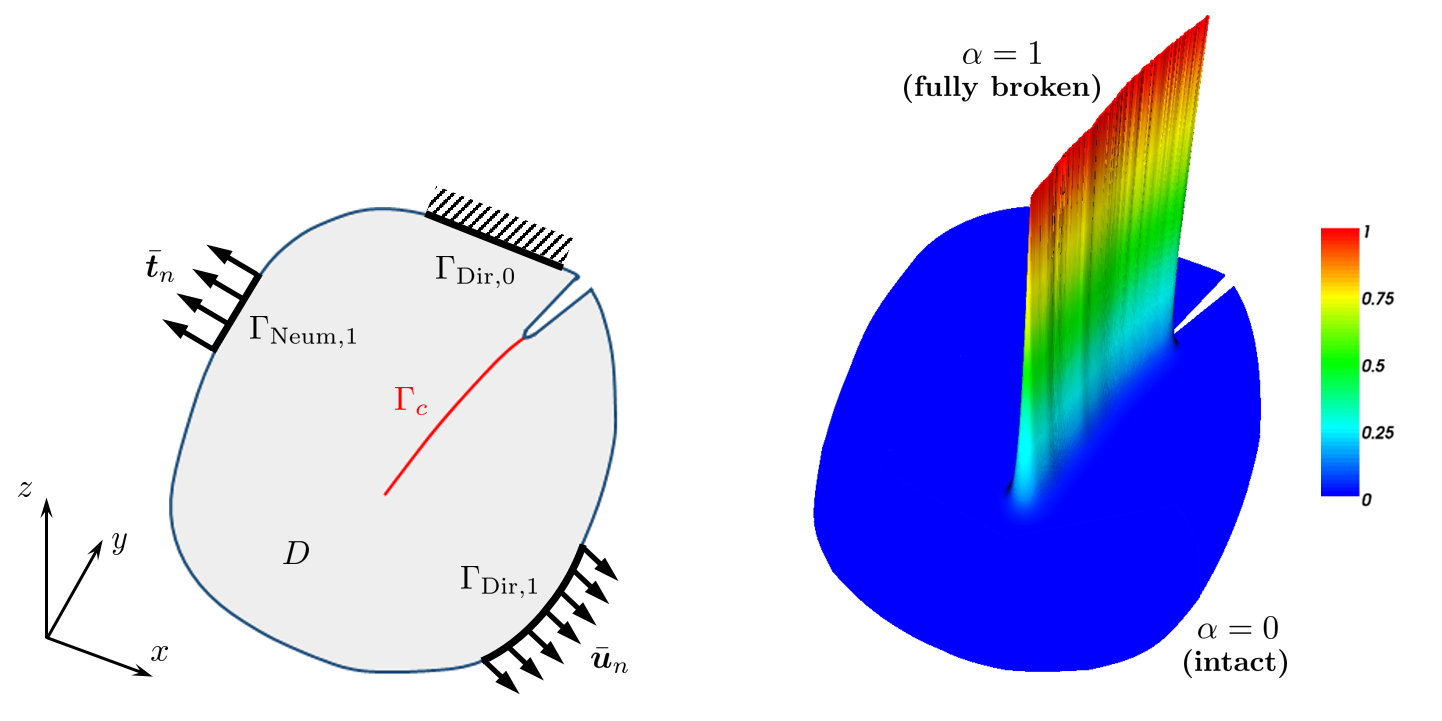}
\end{center}
\caption{Mechanical system and phase-field description of fracture (sketchy in two dimensions) with $\alpha\in C(D;[0,1])$ as the crack phase field.}
\label{fig:PF}
\end{figure}

Let $\Dom\subset\mathbb{R}^m$ ($m=2$ or $3$) be an open and bounded domain representing the configuration of a $d$-dimensional body, and let $\Gamma_{\mathrm{Dir},0},\Gamma_{\mathrm{Dir},1}$ and $\Gamma_{\mathrm{Neum},1}$ be the (non-overlapping) portions of the boundary $\partial \Dom$ of $\Dom$ on which homogeneous Dirichlet, non-homogeneous Dirichlet and Neumann boundary conditions are prescribed, respectively. The material is assumed to be linearly elastic, with the elastic strain energy density function $\Psi(\vek{\varepsilon})$, where $\vek{\varepsilon}:= \vek{\varepsilon}(\vek{u})$ is the infinitesimal strain, $\vek{\varepsilon} = \nabla_s \vek{u}= \tfrac{1}{2}(\nabla \vek{u} + (\nabla \vek{u})^{\ops{T}})$, and $\bm u$ is the displacement. In the isotropic case considered here $\Psi(\vek{\varepsilon})=\frac{1}{2}\lambda\,(\MR{tr}\,\vek{\varepsilon})^2+\mu\,\MR{tr}\,(\vek{\varepsilon}\cdot\vek{\varepsilon})$, with $\lambda$ and $\mu$ as the Lam\'{e} constants. Also, let $G_c$ be the material fracture toughness or critical energy release rate. We consider a quasi-static loading process with the discrete pseudo-time step parameter $n=1,2,...$, such that the displacement $\bar{\bm u}_n$ and traction $\bar{\bm t}_n$ loading data are prescribed on the corresponding parts of the boundary. Finally, let $\Gamma_c\subset D$ be the crack surface that is evolving during the process, see the left plot in Figure \ref{fig:PF}.

For the mechanical system at hand, the variational approach to brittle fracture in \cite{Francfort1998} relies on the energy functional
\begin{equation}
\En(\bm u,\Gamma_c) = 
\int_{D\setminus\Gamma_c} 
\Psi(\bm\varepsilon(\bm u)) \, \mathrm{d}{\bm x}
+ G_c\,\mathbb{H}^{m-1}(\Gamma_c)
-\int_{\Gamma_{\mathrm{Neum},1} } \bar{\bm t}_n \cdot \bm u \, \mathrm{d}s,
\label{VAF}
\end{equation}
and the related minimization problem at each $n\geq 1$. In \eqref{VAF}, $\vek{u}:\Dom\setminus\Gamma_c\to \B{R}^d$, $d=1,2,$ or $3$, such that $\vek{u}=\vek{0}$ on $\Gamma_{\mathrm{Dir},0}$ 
and $\vek{u}=\bar{\vek{u}}_n$ on $\Gamma_{\mathrm{Dir},1}$ is the displacement field, $\vek{\varepsilon} := \vek{\varepsilon}(\vek{u}):\Dom\setminus\Gamma_c\to \B{R}^d$  is the strain field $\Gamma_c$ is the crack set, and $\B{H}^p$ is the so-called $p$-dimensional Hausdorff measure of $\Gamma_c$. The first term in (\ref{VAF}) represents the elastic energy stored in the body, and the second one the fracture surface energy dissipated within the fracture process. In simple terms, $\mathbb{H}^1(\Gamma_c)$ and $\mathbb{H}^2(\Gamma_c)$ are the length and the surface area of $\Gamma_c$ when $d=2$ and $3$, respectively\footnote{Assuming $G_c$ to be non-constant in $D$, a more general representation of the fracture energy term in \eqref{VAF} reads $\displaystyle\int_{\Gamma_c}G_c(\x)\mathbb{H}^{m-1}(\mathrm{d}\bm x)$, see also Section \ref{secProb_gen}.}.

The regularization of \eqref{VAF} {\em \`a la} Bourdin-Francfort-Marigo \cite{Bourdin2000,Bourdin2007a,Bourdin2007b,Bourdin2008}, which is the basis for a variety of fracture phase-field formulations, reads as follows:
\begin{equation}
\En(\bm u,\alpha) = 
\int_D
\tns{g}(\alpha)\Psi(\bm\varepsilon(\bm u)) \, \mathrm{d}{\bm x}
+ \frac{G_c}{c_{\tns w}} \int_D
\left(\frac{\tns{w}(\alpha)}{\ell}+\ell|\nabla\alpha|^2\right)
\mathrm{d}{\bm x}
-\int_{\Gamma_{\mathrm{Neum},1} } \bar{\bm t}_n \cdot \bm u \, \mathrm{d}s,
\label{RegVAF}
\end{equation}
with $\bm u:D\rightarrow\mathbb{R}^d$ and $\alpha:D\rightarrow[0,1]$ standing for the smeared counterparts of the discontinuous displacement and the crack set in \eqref{VAF}. The phase field variable $\alpha$ takes the value $1$ on $\Gamma_c$, decays smoothly to $0$ in a subset of $D\backslash\Gamma_c$ and then 
vanishes in the rest of the domain, as sketched in the right plot in Figure \ref{fig:PF}. With this definition, the limits $\alpha=1$ and $\alpha=0$ represent the fully broken and the intact (undamaged) material phases, respectively, whereas the intermediate range $\alpha\in(0,1)$ mimics the transition zone between them. The function $\tns g$ is responsible for the material stiffness degradation. The function $\tns w$ defines the decaying profile of $\alpha$, whereas the parameter $0<\ell\ll\mathrm{diam}(D)$ controls the thickness of the localization zone of $\alpha$, i.e.\ of the transition zone between the two material states.
  
The specific choice of the functions $\tns g$ and $\tns w$ in \eqref{RegVAF} establishes the rigorous link between \eqref{VAF} and (\ref{RegVAF}) when $\ell\rightarrow 0$ via the notion of $\Upgamma$-convergence, see e.g.\ Braides \cite{Braides1998}, Chambolle \cite{Chambolle2004}, also giving a meaning to the induced constant $c_{\tns w}$. Thus, $\tns g$ is a continuous monotonic function that fulfills the properties: $\tns g(0)=1$, $\tns g(1)=0$, $\tns g^\prime(1)=0$ and $\tns g^\prime(\alpha)<0$ for $\alpha\in[0,1)$, see e.g.\ Pham et al.\ \cite{Pham2011}. The quadratic polynomial $\tns g(\alpha):=(1-\alpha)^2$ is the simplest choice. The function $\tns w$, also called the local part of the dissipated fracture energy density function \cite{Pham2011}, is continuous and monotonic such that $\tns w(0)=0$, $\tns w(1)=1$ and $\tns w^\prime(\alpha)\geq 0$ for $\alpha\in[0,1]$. The constant $c_{\tns w}:=4\int_0^1\sqrt{\tns w(t)}\,\mathrm{d}t$ is a normalization constant in the sense of $\Upgamma$-convergence. The two suitable candidates for $\tns w$ which are widely adopted read $\tns w(\alpha)=\alpha$ and $\alpha^2$, such that $c_{\tns w}=\frac{8}{3}$ and $2$, respectively.

The combinations of formulation (\ref{RegVAF}) with the aforementioned choices for $\tns g$ and $\tns w$ are typically termed the $\mathtt{AT}$-1 and $\mathtt{AT}$-2 models, see Table \ref{AT}. $\mathtt{AT}$ stands for {\em Ambrosio-Tortorelli} and the corresponding type of regularization, see \cite{AT1990}. The main difference between the two models is that $\mathtt{AT}$-1 leads to the existence of an elastic stage before the onset of fracture, whereas using $\mathtt{AT}$-2 the phase-field starts to evolve as soon as the material is loaded, see e.g.\ \cite{Amor2009,Pham2011,Marigo2016} for a more detailed explanation. Other representations for $\tns g$ and $\tns w$ are available in the literature, see e.g.\ \cite{Borden2014,Kuhn2015,Sargado2018,Burke2013}.

\begin{table}[h]
\caption{\em Ingredients of formulation (\ref{RegVAF}).}
\centering
      \begin{tabular}{c|c|c}
      \hline
        $\tns g$   &  $\tns w$ &  name  \\
	\hline
        $(1-\alpha)^2$ &  \begin{tabular}{c} $\alpha$ \\   $\alpha^2$ \end{tabular}  
				     &  \begin{tabular}{c} $\mathtt{AT}$-1 model \\ $\mathtt{AT}$-2 model \end{tabular} \\
	\hline
      \end{tabular}
\label{AT}
\end{table}

With $\En$ defined by (\ref{RegVAF}), the sought solution at a given loading step $n\geq 1$ is given by
\begin{equation}
(\Vu,\alpha) =\argmin \, \{ \En(\Vv,\beta): \; \Vv\in \sU_{n}, \beta\in \sP_{n} \}.
\label{argmin0}
\end{equation}
Here
\begin{equation}   \label{eq:Un}
\sU_{n}:=\{\Vu\in \Hp^1(\Dom;\B{R}^d): \; 
\Vu=\vek{0} \; \MR{on} \; \Gamma_{\MR{Dir},0},
\; \Vu=\bar{\Vu}_n\; \MR{on} \; \Gamma_{\MR{Dir},1} \}
\end{equation}
is the kinematically admissible affine displacement space satisfying the non-homogeneous Dirichlet condition at load step $n$ with $\Hp^1(\Dom;\B{R}^d)$ as the usual Sobolev space of functions with values in $\B{R}^d$, and
\begin{equation}  \label{eq:Pn}
\sP_{n}:=\{ \alpha\in \sP: \; \alpha\geq\alpha_{n-1} \; \MR{in} \; \Dom \} 
    \subset \sP := \Hp^1(\Dom;\B{R})
\end{equation}
is the admissible convex subset for $\alpha$ at time $n$ with $\alpha_{n-1}$ known from the previous loading step
in the Sobolev space of phase fields $\sP$. The condition $\alpha\geq\alpha_{n-1}$ in $\Dom$ is used to enforce the {\em irreversibility} of the crack phase field evolution. It is the backward difference quotient form of $\dot{\alpha}\geq0$ in $D$.

The necessary optimality conditions for $(\bm u,\alpha)\in \sU_{n}\times \sP_{n}$ at every loading step $n\geq1$ read as follows:
\begin{equation}
\left\{
\begin{tabular}{l}
$\En_{\Vu}(\Vu,\alpha;\Vv) = \dual{\Gdiff_{\Vu} \En(\Vu,\alpha)}{\Vv} = 0 \quad\forall \Vv\in\C{U}$,   \\[0.2cm]
$\En_\alpha(\Vu,\alpha;\beta-\alpha) = \dual{\Gdiff_{\alpha} \En(\Vu,\alpha)}{\beta-\alpha}\geq 0\quad\forall \beta\in\C{P}_n$,
\end{tabular}
\right.
\label{Weak}
\end{equation}
where $\En_{\Vu}$ resp.\ $\Gdiff_{\Vu} \En$ and $\En_{\alpha}$ resp.\ $\Gdiff_{\alpha} \En$ denote 
the partial G\^ateaux derivatives (first variations) of the energetic functional $\En$ w.r.t.\ $\Vu$ and
$\alpha$, and $\dual{\cdot}{\cdot}$ is an appropriate duality pairing. The displacement test space in \eqref{Weak} is
\begin{equation}  \label{eq:U-def}
\sU := \Hp^1_{\Gamma}(\Dom;\B{R}^d):= \{\Vv\in \Hp^1(\Dom;\B{R}^d): \; 
          \Vv=\vek{0} \; \MR{on} \; \Gamma_{\MR{Dir},0} \cup \Gamma_{\MR{Dir},1} \},
\end{equation}
the displacement fields with homogeneous boundary conditions. As noted in the introduction, conditions \eqref{Weak} characterize in general a local minimum of $\En$ (or even only a local stationary point).

%%%%%%%%%%%%%%%%%%%%%
\subsection{Model example: anti-plane shear test}
\label{CaseStudy}

\begin{figure}[h]
\begin{center}
\includegraphics[width=1.0\textwidth]{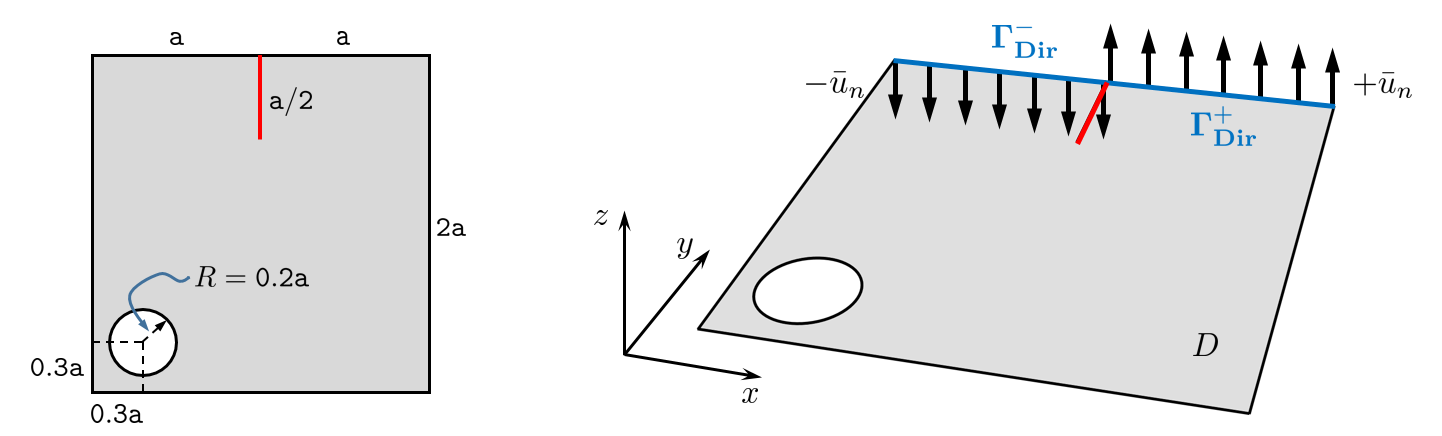}
\end{center}
\caption{Geometry and loading setup for the anti-plane shear experiment.}
\label{fig:setup}
\end{figure}

Following \cite{Artina2015}, we consider a two-dimensional rectangular domain $(0,\mathtt{2a})\times(0,\mathtt{2a})$ containing a slit along $\{\mathtt{a}\}\times(\mathtt{3a/2},\mathtt{2a})$ and a circular hole with center $(\mathtt{0.3a},\mathtt{0.3a})$ of radius $R=\mathtt{0.2a}$. The incremental anti-plane displacement $\mp \bar{u}_n$, $n\geq1$ is applied on $\Gamma_\mathrm{Dir}^{-}:=(0,\mathtt{a})\times\{\mathtt{2a}\}$ and $\Gamma_\mathrm{Dir}^{+}:=(\mathtt{a},\mathtt{2a})\times\{\mathtt{2a}\}$, respectively, see Figure \ref{fig:setup}.

The energy functional used in the incremental minimization problem in this case reads:
\begin{equation}
\En(u,\alpha) = \frac{1}{2} \int_D
 (1-\alpha)^2\mu\,|\nabla u|^2 \, \mathrm{d}{\bm x}
+  \frac{G_c}{c_{\sf w}} \int_D
\left(\frac{\tns w(\alpha)}{\ell}+\ell|\nabla\alpha|^2\right)
\mathrm{d}{\bm x}
+\frac{\gamma}{2}\int_D{\left\langle\alpha-\alpha_{n-1}\right\rangle}_{-}^2
\mathrm{d}{\bm x}.
\label{PF}
\end{equation}
The first term in (\ref{PF}) is the corresponding one from the standard formulation in (\ref{RegVAF}) adapted to the anti-plane shear situation: we assume $\bm u=(0,0,u_z)$ and $u_z = u:\Dom\subset\B{R}^2\to\B{R}$, such that $\vek{\varepsilon} = \nabla u: \Dom\to\B{R}^2$ with $\nabla:=[\dd x,\dd y]^{\ops{T}}$. The last term is a penalty term which enforces the irreversibility constraint $\alpha\geq\alpha_{n-1}$ via penalization with $\langle y \rangle_{-}:=\min(0,y)$ and $\gamma\in\mathbb{R}_{+}$ as the penalty parameter, see \cite{Gerasimov2019}.

Upon the incorporation of the penalty term, the necessary conditions (\ref{Weak}) turn into a system of equalities, reading
\begin{equation}
\left\{
\begin{tabular}{l}
$\En_u(u,\alpha;v) =\dual{\Gdiff_{u} \En(u,\alpha)}{v}=0 \quad \forall \; {v}\in \mathcal{U}$,   \\[0.2cm]
$\En_\alpha({u},\alpha;\beta)=\dual{\Gdiff_{\alpha} \En(u,\alpha)}{\beta}=0 \quad \forall \; \beta\in \mathcal{P}$,
\end{tabular}
\right.
\label{Weak_aps}
\end{equation}
where
\begin{align}   \label{eq:Weak_aps-u}
\En_u(u,\alpha;v) =& 
\dual{\Gdiff_{u} \En(u,\alpha)}{v} :=
\int_{\Dom} 
(1-\alpha(\vek{x}))^2\mu\,\nabla u(\vek{x})\cdot\nabla v(\vek{x}) \,\di\vek{x},
\\   \label{eq:Weak_aps-a}
\En_\alpha(u,\alpha;\beta) =& 
\dual{\Gdiff_{\alpha} \En(u,\alpha)}{\beta} := 
 -\int_{\Dom} \mu\, (1-\alpha(\vek{x}))\,
|\nabla u(\vek{x})|^2\,\beta(\vek{x}) \,\di \vek{x}
\\  \nonumber
 &+ \frac{G_c}{c_{\tns w}}
\int_{\Dom}\left( \frac{1}{\ell}{\tns w}^\prime(\alpha(\vek{x})) \beta(\vek{x})
+ 2\,\ell\,\nabla \alpha(\vek{x})
\cdot \nabla \beta(\vek{x}) \right)
\, \di \vek{x}
\\  \nonumber
&+\gamma\; 
\int_{\Dom}\langle \alpha(\vek{x})-\alpha_{n-1}(\vek{x}) \rangle_{-}\,\beta(\vek{x}) \,\di\vek{x}.
\end{align}

% Algorithm

\begin{table}[h]
\small
\centering
{
\begin{tabular}{l}
\hline \\
{\bf Input:} loading data $\pm\bar{u}_n$ on $\Gamma_\mathrm{Dir}^\pm$, and \\ 

{\color{white}\bf Input:} solution $(u_{n-1},\alpha_{n-1})$ from step $n-1$. \\ [0.1cm]

\quad\quad Initialization, $k=0$:\\
\quad\quad\quad 1.\; set $(u^{(0)},\alpha^{(0)}):=(u_{n-1},\alpha_{n-1})$.\\ [0.1cm]

\quad\quad Staggered iteration $k\geq1$:\\

\quad\quad\quad 2.\; given $u^{(k-1)}$, solve $\displaystyle \En_\alpha(u^{(k-1)},\alpha;\beta)=0$ $\forall\beta$
for $\alpha$, set $\alpha=:\alpha^{(k)}$,\\

\quad\quad\quad 3.\; given $\alpha^{(k)}$, solve $\En_u(u,\alpha^{(k)};v)=0$ $\forall v$ for $u$, set $u=:u^{(k)}$,\\

\quad\quad\quad 4.\; for the obtained pair $(u^{(k)},\alpha^{(k)})$, check \\

\quad\quad\quad\quad\quad\quad\quad
$\mathrm{Res}_\mathrm{Stag}^{(k)}:=|\displaystyle \En_\alpha(u^{(k)},\alpha^{(k)};\beta)|
\leq\texttt{TOL}_\mathrm{Stag} \; \forall\beta$, \\

\quad\quad\quad 5.\; if fulfilled, set $(u^{(k)},\alpha^{(k)})=:(u_n,\alpha_n)$ and stop; \\

\quad\quad\quad {\color{white}5.}\; else $k+1\rightarrow k$. \\ [0.1cm]

{\bf Output:} solution $(u_n,\alpha_n)$. \\ \\
\hline
\end{tabular}
}
\caption{Staggered iterative solution algorithm for \eqref{Weak_aps} at loading step $n\geq 1$.}
\label{TableStag}
\end{table}

The staggered solution algorithm for the system in \eqref{Weak_aps} implies alternately fixing $u$ and $\alpha$ and solving the corresponding equations until convergence. The algorithm is sketched in Table \ref{TableStag}. Note that the phase-field evolution equation $\En_\alpha=\Gdiff_\alpha \En=0$ is non-linear due to the Macaulay brackets term $\langle\cdot\rangle_{-}$. Therefore, a Newton-Raphson procedure is used to iteratively compute $\alpha^{(k)}$ with $\alpha^{(k-1)}$ taken as the initial guesses and $\mathtt{TOL}_\mathrm{NR}$ as the tolerance for the corresponding residual. 

\begin{figure}[h]
\begin{center}
\includegraphics[width=0.8\textwidth]{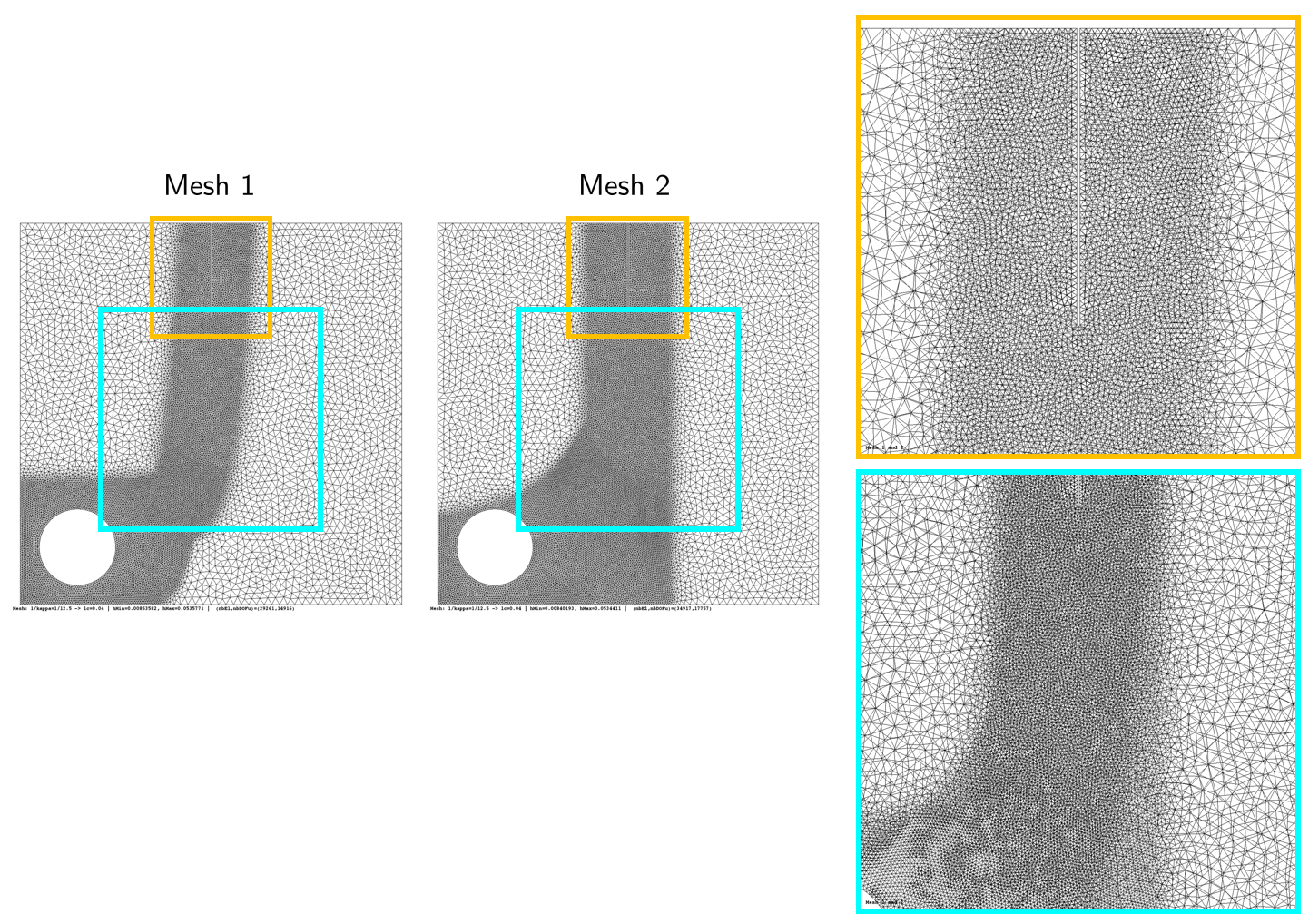}
\end{center}
\caption{Finite element meshes (on the left) and their overlapping in the marked regions (on the right); the latter is to illustrate that one mesh can be viewed as a perturbation of the other one.}
\label{fig:meshes1and2}
\end{figure}

\begin{table}[h]
    \centering
    \begin{tabular}{l||c|c|c}
    \hline
    Crack path type & \text{Type 1} & \text{Type 2} & \text{Type 3} \\
    \hline
    Crack path (schematically)
    & \includegraphics[width=0.05\textwidth]{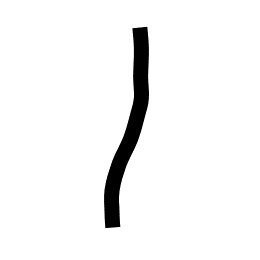} 
    & \includegraphics[width=0.05\textwidth]{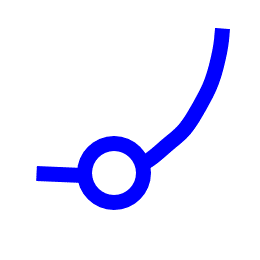} 
    & \includegraphics[width=0.05\textwidth]{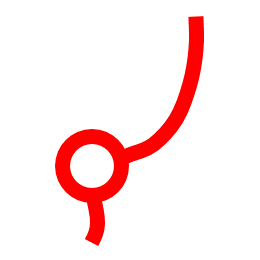} \\
    \hline
    \end{tabular}
    \caption{Crack paths classification.}
    \label{tab:CrackClass}
\end{table}

In the computations, we set $\mathtt{a}=1$, $\mu=1$, $\ell=2\mathtt{a}/50=0.04$, $G_c=1$. In what follows, we choose the quadratic local function $\tns w(\alpha)=\alpha^2$ such that $c_{\tns w}=2$. Also, we set $\gamma=\frac{G_c}{\ell}\left( \frac{1}{\mathtt{TOL}_\mathrm{ir}^2}-1 \right)$, where $\mathtt{TOL}_\mathrm{ir}=0.01$. As argued in \cite{Gerasimov2019}, this choice of $\gamma$ provides a sufficiently accurate enforcement of $\alpha\geq\alpha_{n-1}$. The applied displacement is given by $\bar{u}_n=n\Updelta\bar{u}$, $n=1,...,\frac{3}{2\Updelta\bar{u}}$, with $\Updelta\bar{u}$ as the loading
increment. The deterministic results will be presented for $\Updelta\bar{u}\in\{0.01,0.1\}$ in order to evaluate the impact of the increment size on the ability to trigger solution non-uniqueness. The error tolerances are prescribed as $\mathtt{TOL}_\mathrm{NR}:=10^{-6}$ and $\mathtt{TOL}_\mathrm{Stag}:=10^{-4}$.

\begin{figure}[h]
\begin{center}
\includegraphics[width=1.0\textwidth]{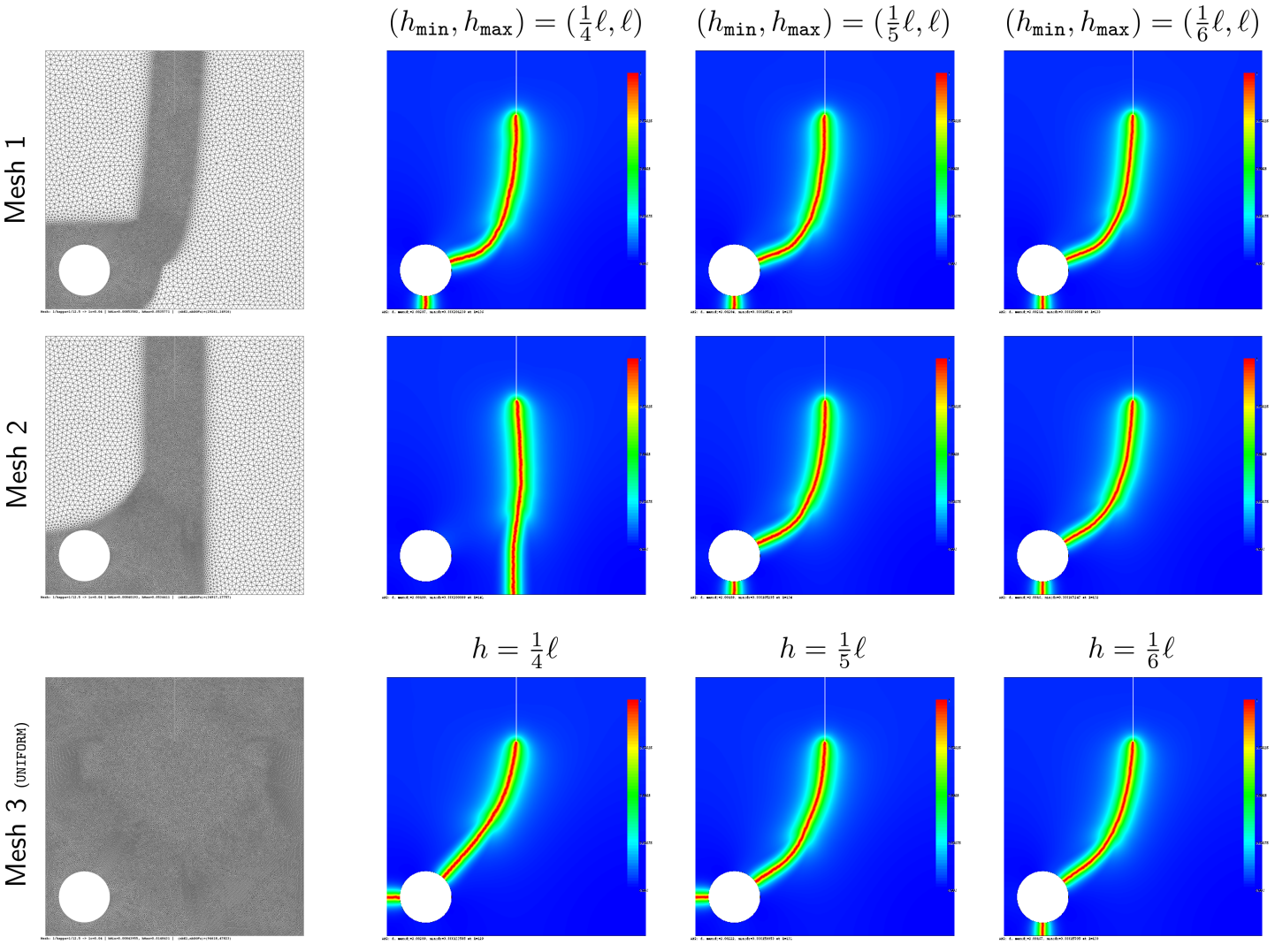}
\end{center}
\caption{Computational results on three different mesh types with varying minimum mesh size for the loading increment $\Updelta\bar{u}=0.01$. Three different fracture mechanisms can be observed.}
\label{fig:DetResAlphIncr001}
\end{figure}

In our simulations we employ the numerical package FreeFem++ \cite{FreeFem}. Both the displacement field $u$ and the crack phase-field $\alpha$ are approximated using $P_1$-triangles. We construct three types of finite element meshes. $\sf{Mesh\;1}$ and $\sf{Mesh\;2}$ are pre-adapted meshes which are refined in the region of $D$ where crack propagation is expected, see Figure \ref{fig:meshes1and2}. They differ only by the shape of the refined region, but have identical mesh size characteristics $(h_\mathtt{min},h_\mathtt{max})$. Here, $h_\mathtt{min}$ and $h_\mathtt{max}$ stand for the mesh size inside and outside of the refined region, respectively. The left plot in Figure \ref{fig:meshes1and2} depicts the meshes with $(h_\mathtt{min},h_\mathtt{max}):=(\frac{1}{4}\ell,\ell)$. Notice that the right plot of the figure aims at illustrating the perturbation character of the considered meshes (that is, one of them can be viewed as a perturbation of the other one), as we assume such perturbation is needed for capturing non-unique solutions. Additionally to $\sf{Mesh\;1}$ and $\sf{Mesh\;2}$, we also consider a uniform mesh whose mesh size $h$ is set to $h_\mathtt{min}$ used in the corresponding pre-adapted cases. This uniformly fine mesh denoted as $\sf{Mesh\;3}$ can be treated as the reference one in the sense of the solution discretization error, as quantities like, e.g. the elastic energy, the fracture energy, as well as the total energy are computed most accurately on this mesh. In the following deterministic computations, we also vary the minimum mesh size $h_\mathtt{min}$, namely, we set $h_\mathtt{min}\in\{\frac{1}{4}\ell,\frac{1}{5}\ell,\frac{1}{6}\ell\}$. For a given $\sf{Mesh\;n}$, ${\sf n}=1,2,3$, this can be viewed as another kind of mesh perturbation, and we intend to assess also its effect on the possibility to trigger multiple solutions.

%%%%%%%%%%%%%%%%%%%%%
\subsection{Numerical results}
\label{NumDet}
\begin{figure}[h]
\begin{center}
\includegraphics[width=0.9\textwidth]{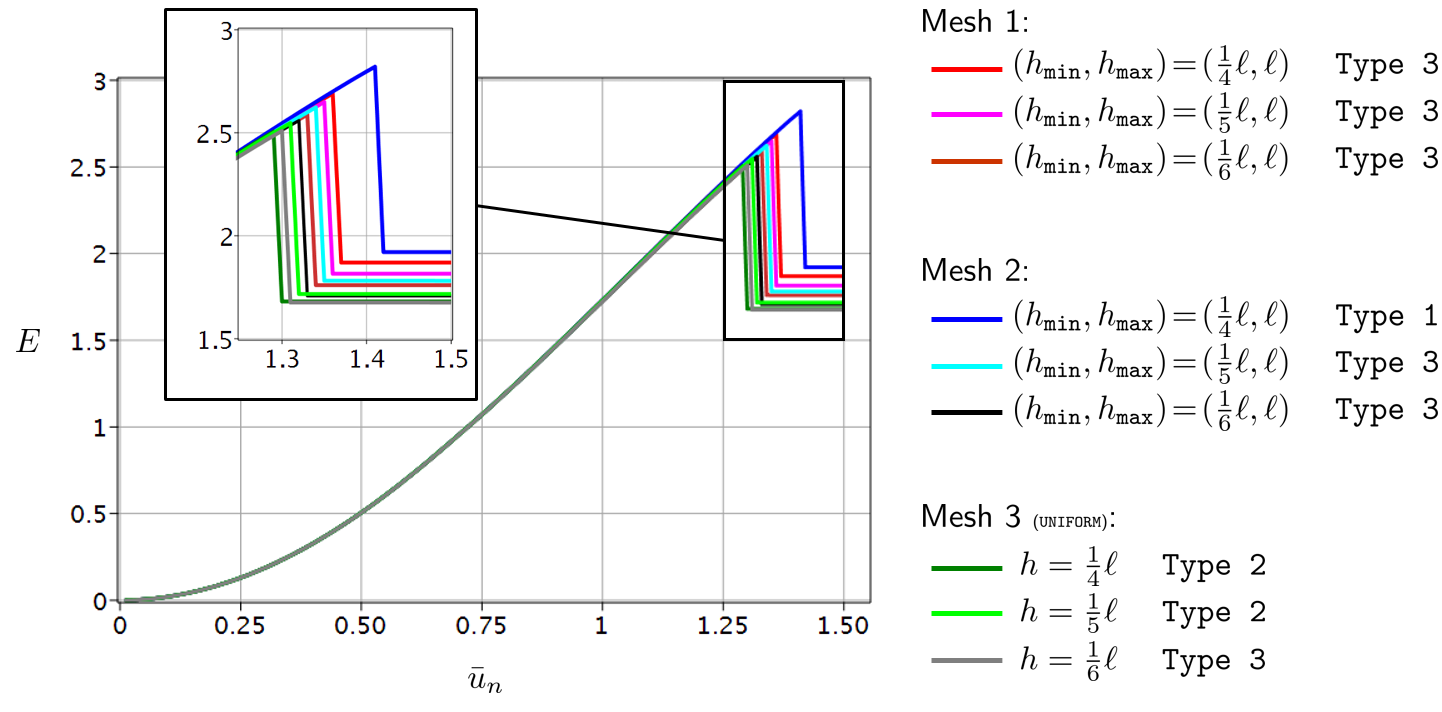}
\end{center}
\caption{Energy-displacement plots for the solutions from Figure \ref{fig:DetResAlphIncr001}.}
\label{fig:DetResEnergIncr001}
\end{figure}

In Figure~\ref{fig:DetResAlphIncr001}, the computational results for the three described types of mesh with varying minimum mesh size and for the loading increment $\Updelta\bar{u}=0.01$ are presented. As expected, both the change of the type of mesh and the change of $h_\mathtt{min}$ may lead to a change in the final crack pattern. Interestingly, additionally to the two fracture mechanisms reported in \cite{Artina2015} (curved crack path which connects the notch and the hole, and then leaves the hole either vertically or horizontally), we also observe a third one which is represented by the (almost) vertical crack that seems attracted by the hole only slightly, yet does not reach it, see the plot for $\sf{Mesh\;2}$ with $(h_\mathtt{min},h_\mathtt{max}):=(\frac{1}{4}\ell,\ell)$ in Figure \ref{fig:DetResAlphIncr001}. In Table \ref{tab:CrackClass}, we assign to each of these crack paths the corresponding type. This classification will be also used in Section \ref{secProb_gen}. Figure \ref{fig:DetResEnergIncr001} depicts the corresponding energy-displacement curves. From the energy plots it may be concluded that the Type 1 (nearly vertical) crack path is not energetically favorable in comparison with Types 2 and 3 (curved paths). Also, the latter ones seem to have almost identical energy levels.

\begin{figure}[h]
\begin{center}
\includegraphics[width=1.0\textwidth]{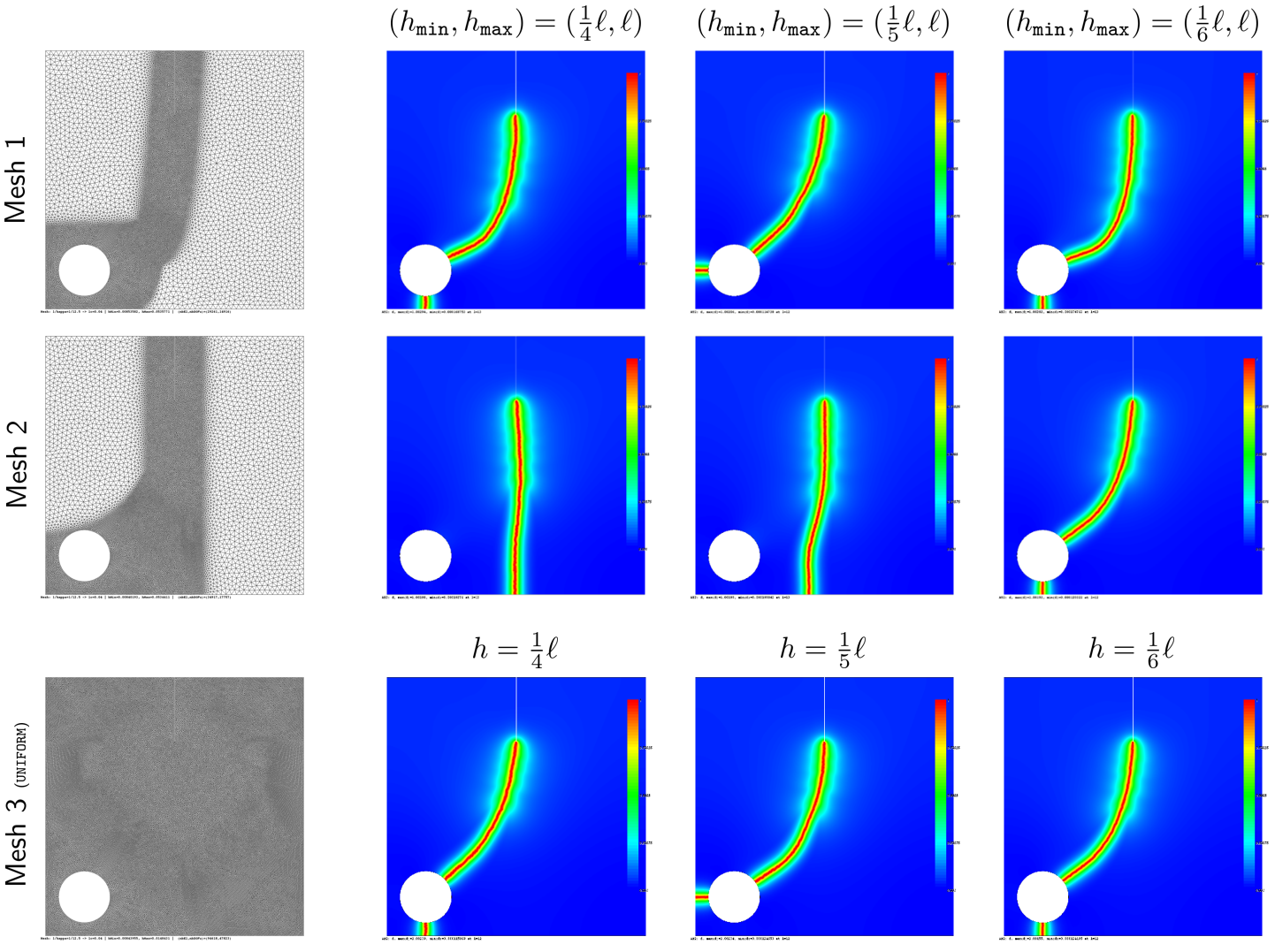}
\end{center}
\caption{Computational results on three different mesh types with varying minimum mesh size for the loading increment $\Updelta\bar{u}=0.1$. Fracture mechanisms similar to those in Figure \ref{fig:DetResAlphIncr001} are observed.}
\label{fig:DetResAlphIncr01}
\end{figure}

The computational time for the numerical experiment with $\Updelta\bar{u}=0.01$, depending on the mesh size, ranges from 6 to 14 hours on a standard desktop machine (Intel(R) Core(TM) i7-3770 OK, CPU 3.5 GHz, RAM 16.0 GB). Therefore, we test also the larger loading increment $\Updelta\bar{u}=0.1$. In Figure \ref{fig:DetResAlphIncr01}, the corresponding results are depicted. Our first observation is that the alteration of the loading increment while keeping the mesh type fixed can trigger solution non-uniqueness, as can be seen by comparing the results for, e.g. $\sf{Mesh\;1}$ with $(h_\mathtt{min},h_\mathtt{max})=(\frac{1}{5}\ell,\ell)$ from Figures \ref{fig:DetResAlphIncr001} and \ref{fig:DetResAlphIncr01}, respectively. Similar observations hold for $\sf{Mesh\;2}$ with $(h_\mathtt{min},h_\mathtt{max})=(\frac{1}{5}\ell,\ell)$, and also for $\sf{Mesh\;3}$ with $h=\frac{1}{4}\ell$ from the corresponding figures. Moreover, with the larger increment and regardless of the mesh type, we are capable to generate exactly the same three types of fracture mechanisms as in the previous case of $\Updelta\bar{u}=0.01$ but with a significantly lower computational effort: the time spent now ranges from 1 to 3 hours. Therefore, in the forthcoming stochastic modeling, which requires numerous realizations, the larger increment will be used.

%%%%%%%%%%%%%%%%%%%%%%%%%%%%%%%%%%%%%%
\section{Stochastic modeling: some informal examples}
\label{secProb_1D}

This section is intended as an informal introduction to the ideas on how to
model minimisation problems with multiple solutions in a relaxed setting, where we
introduce perturbations of the functional and allow random variables as solutions.
The first example deals with the minimisation of an ordinary function, whereas the next two examples address fracture of a 1D bar under tension, first with a sharp crack and then with a phase-field approach.

For these examples only rudimentary notions of probability are required, the details may
be found in Subsection~\ref{sec:random-variables-and-probability-distributions}.  All
we need to know is that when $\tns{x}$ is a random variable (RV), and 
$\varphi(\tns{x})$ some function of this RV, the expectation of $\varphi$ is $\B{E}[\varphi] :=
\int \varphi(\tns{x})\, \di \P$, where $\P$ is the underlying probability measure.
%} 

\subsection{Minimization of a double-well function}\label{sec:introductory-concepts}

For a start, consider the example of the double-well function $\En(x) = x^2(1-x)^2$
defined on the real line $\sR$, which is not convex and has two global minima at the
points $x=0$ and $x=1$ with value $\En(0)=\En(1)=0$, and a local maximum at the point $x=1/2$.

The original minimisation problem of finding $x=\argmin_{y\in\sR} \En(y)$
with the solutions $x=0$ and $x=1$ will now be converted --- relaxed --- to a minimisation
over RVs.  We will perturb the function $\En(\cdot)$ by a 
RV $\rv{q}$ to $\En(\rv{q};\cdot)$ in such a way
that $\En(0;\cdot)$ is the original unperturbed function, and will be
looking for minimisers of the expected value of $\En$ not in $\sR$, 
but in some appropriate space $\C{X}$ of RVs with values in $\sR$:
\[
  \rv{x} = \argmin_{\rv{y}\in\C{X}} \Ens(\rv{y})
          = \argmin_{\rv{y}\in\C{X}} \E{\En(\rv{q};\rv{y})} .
\]
Here the new function is $\Ens(\rv{y})=\E{\En(\rv{q};\rv{y})}$, %\JVc{expectation is not explained} 
and we take as perturbation $\En(\rv{q};\rv{x})= \rv{x}^2\cdot(1-\rv{x})^2 + \rv{q}\cdot \rv{x}$. More generally, we are interested in 
\begin{equation}  \label{eq:per-0D-prob-eta}
\rv{x}(\eta) = \argmin_{\rv{y}\in\C{X}} \E{\En(\eta\cdot\rv{q};\rv{y})},
\end{equation}
for $\eta>0$ and in the limiting behaviour $\lim_{\eta\to +0} \rv{x}(\eta)$. 

We first consider
the minimisation over $\C{X}$ without perturbation, i.e.\ $\eta=0$. 
As the minimum of the function is still zero,
it is not difficult to see that two solutions are the two --- not really random ---
variables $\rv{x}\equiv 0$ and $\rv{x}\equiv 1$.
%which are constant identically equal to $0$ resp.\ $1$.  
In fact, any RV $\rv{x}$ which takes only the value $0$ with probability $\P(\rv{x}=0) = p$ and 
the value $1$ with probability $\P(\rv{x}=0) = 1-p$ for any 
$p \in [0,1]$ is a minimiser. As explained in
Subsection~\ref{sec:random-variables-and-probability-distributions}, for a fixed $p \in [0,1]$ all such RVs are considered as equivalent.
Therefore, the probability distribution $\mv_{\rv{x}}$ of \emph{any} minimiser 
$\rv{x}$ can be described as the 
convex combination of two Dirac point measures $\updelta_x$ located at the global minima
$x=0$ and $x=1$, i.e.\ 
\[ 
\mv_{\rv{x}} = p\, \updelta_0 + (1-p)\, \updelta_1 \;\text{ for any } p\in[0,1] .
\]

Let us consider now the perturbed (still non-convex) problem \eqref{eq:per-0D-prob-eta} with a bounded perturbing RV $\rv{q}$ and sufficiently small $0\le\eta\ll 1$. The perturbed function still has two local minima and in between a local maximum.  Thus, for a positive realisation of the RV $\rv{q}$ the global
minimum is still close to $x=0$, and for a negative realisation of the RV $\rv{q}$ 
the global minimum is still close to $x=1$, continuously dependent on $\eta$.
Setting $\bar{p} := \P(\rv{q} > 0)$,
it is not difficult to see that the stochastic solution  for $\eta\to +0$
is any RV $\rv{x}$ which takes only the values $\rv{x}=0$ with probability 
$\P(\rv{x}=0)=\bar{p}$ and $\rv{x}=1$ with probability $\P(\rv{x}=1)=1-\bar{p}$.
Here it was tacitly assumed that the RV $\rv{q}$ takes the value $0$ with vanishing probability: $\P(\rv{q}=0) = 0$.  
And as RVs which have the same distribution are considered
equivalent, cf.\ Subsection~\ref{sec:random-variables-and-probability-distributions},
this now unique abstract 
%--- as $\bar{p}$ is a fixed number dependent on the RV $\rv{q}$ --- 
RV has the limiting distribution
\[
  \mv_{\rv{x}} = \bar{p}\, \updelta_0 + (1-\bar{p})\, \updelta_1,
\]
where $\bar{p}$ is a fixed number dependent on the RV $\rv{q}$. Summing up, the limiting solution for the relaxed (stochastic) global minimization problem in this case is unique, and it depends on the perturbation $\rv{q}$.

Thus far we considered the global minimization of $\En(x)$. Let us now consider the relaxation of the corresponding deterministic stationarity condition 
$\di \En(x)/\di x = 0$ with solutions at $x = 0, 1/2$, and $1$. The Euler-Lagrange equation for the solution $\rv{x}(\eta)$ is (cf.\ Subsection~\ref{sec:stochastic-formulations-of-variational-problems})
\begin{equation} \label{Euler-Lagrange}
%\ipd{\Gdiff_{\uu}\rv{J}(\rv{x})}{\rv{y}} := 
   \E{\dd_x \En(\eta\cdot\rv{q};\rv{x}(\eta))\cdot\rv{y}} = 0 
        \quad \forall \rv{y}\in\C{X} .
\end{equation} 
With a bounded perturbing RV $\rv{q}$ and sufficiently small $0\le\eta\ll 1$, \eqref{Euler-Lagrange} has three solutions --- stationary points of $\En$ --- for any
realisation of the RV $\rv{q}$. Thus the solutions, which depend continuously on $\eta$, are still going to be close to 
$x = 0, 1/2$, and $1$, which are the limiting solutions for $\eta\to +0$. Hence, any RV $\rv{x}$ which only takes the values  $x = 0, 1/2$, and $1$ with arbitrary probabilities $\P(\rv{x}=0)=p_0$,
$\P(\rv{x}=1)=p_1$ (with $p_0+p_1 \le 1$), and $\P(\rv{x}=1/2)=(1-p_0-p_1)$ is a limiting solution, and the probability distribution $\mv_{\rv{x}}$ of \emph{any} such 
limiting stationary solution $\rv{x}$ is
\[ 
\mv_{\rv{x}} = p_0\, \updelta_0 + (1- p_0 - p_1)\, \updelta_{1/2}
 +  p_1\, \updelta_1 .
\]
The Euler-Lagrange equation can neither distinguish between maximum or minimum, nor
between local and global minimum.  Therefore the limiting solution for the relaxed (stochastic) stationarity problem embodied by the Euler-Lagrange equation in this case is not unique, and it does not depend on $\rv{q}$.

\subsection{Fracture of a 1D bar with a sharp crack approach}
\label{sec:1D_griffith}

We next consider fracture of a 1D bar. This example is inspired by the study in \cite{Francfort1999_1D} and similar localisation studies for gradient-based plasticity and damage models \cite{JiZeVo2010SGP,Rokos2015,Jirasek2015,Jirasek2013locgradplas}.

Consider a bar of length $L$, with varying cross-section $A(x)$, fixed at one end and subjected to an increasing applied displacement at the other end. The energy functional of brittle 
fracture \eqref{VAF} is in this case with $\Dom=[0,L]$:
\begin{align*}
\En(u,\Gamma_c) = \frac{1}{2}\int_{D \setminus \Gamma_c}  Y A(x) (u'(x))^2 \D{x} + 
   \int_{\Gamma_c} G_cA(x) \Dm{\Hm^0}{x} ,
\end{align*}
where the Young's modulus $Y$ and the critical energy release rate $G_c$ are constant. $G_cA(x)$ is the fracture
energy for a crack at $x$, and $\Hm^{0}(\Gamma_c)$ is the Hausdorff measure of the crack,
a discrete measure equal to the number of crack points $\# \Gamma_c$. In the numerical tests we set $L=6$. Note that although dimensionally $G_cA(x)$ is a dissipated energy, in this 1D example where fracture occurs at a point it can also be considered a dissipation density, and we will refer to it accordingly. In this example, due to the difficulty in dealing with the stationarity condition with respect to the unknown crack set $\Gamma_c$, we only consider the global minimization of the energy and its relaxed (stochastic) formulation.

The bar does not crack until the elastic energy 
$\frac{1}{2}\int_\sY  Y A(x) (u'(x))^2 \D{x}$ reaches the minimum dissipation $\min_{x\in\sY} G_cA(x)$ with a single crack. A failure with multiple cracks cannot occur in the formulation with global minimisation of the energy because a higher dissipation would be required. Note that there is no need for a real computation, as the location of the crack 
$x_c=\argmin_{x\in\sY} G_cA(x)$ depends only on the location of the minimum dissipation density.

%\begin{figure}[htb]
\begin{figure}[h!]
 \centering
 \subfloat[]{\includegraphics[width=0.45\textwidth]{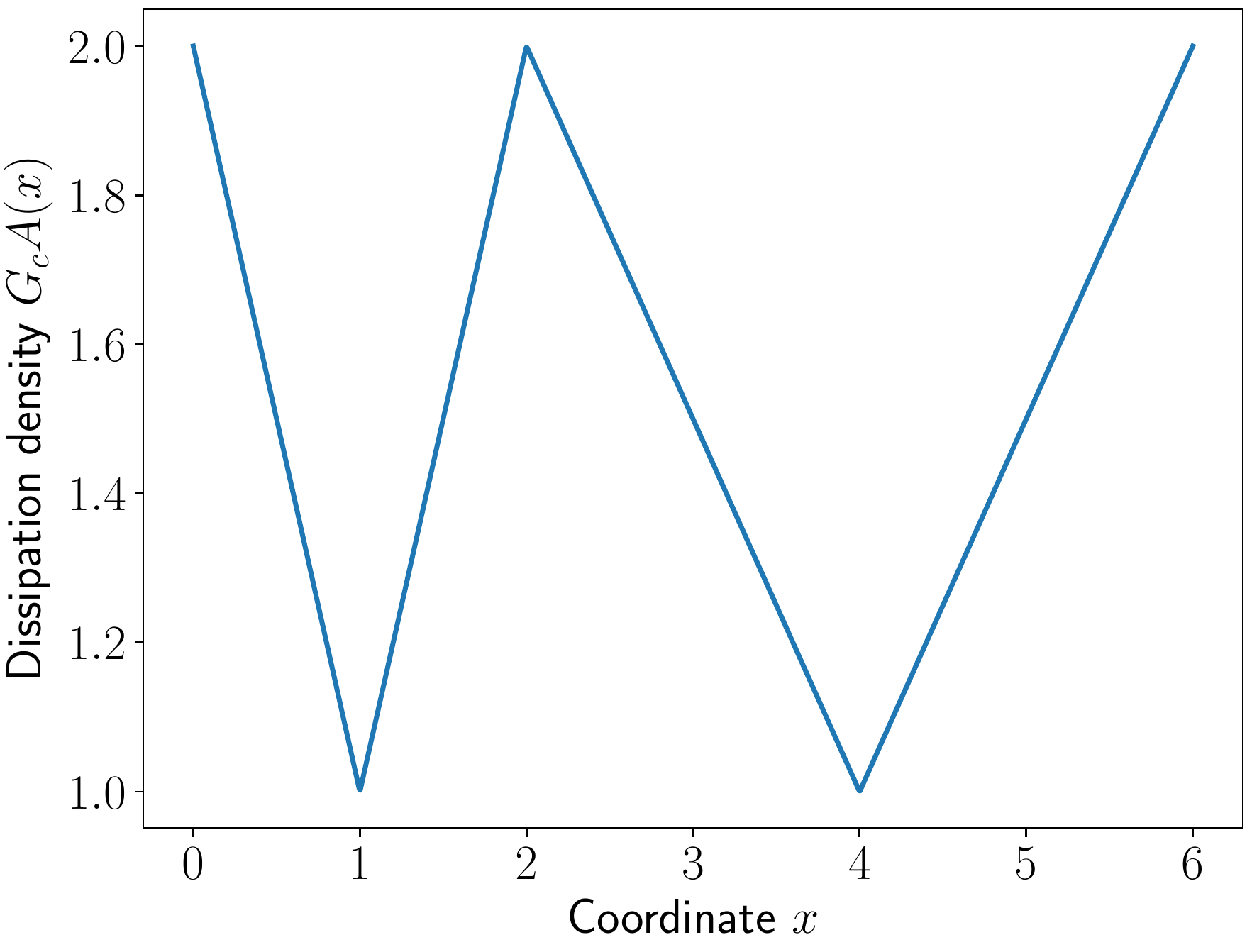}}
 \subfloat[]{\includegraphics[width=0.45\textwidth]{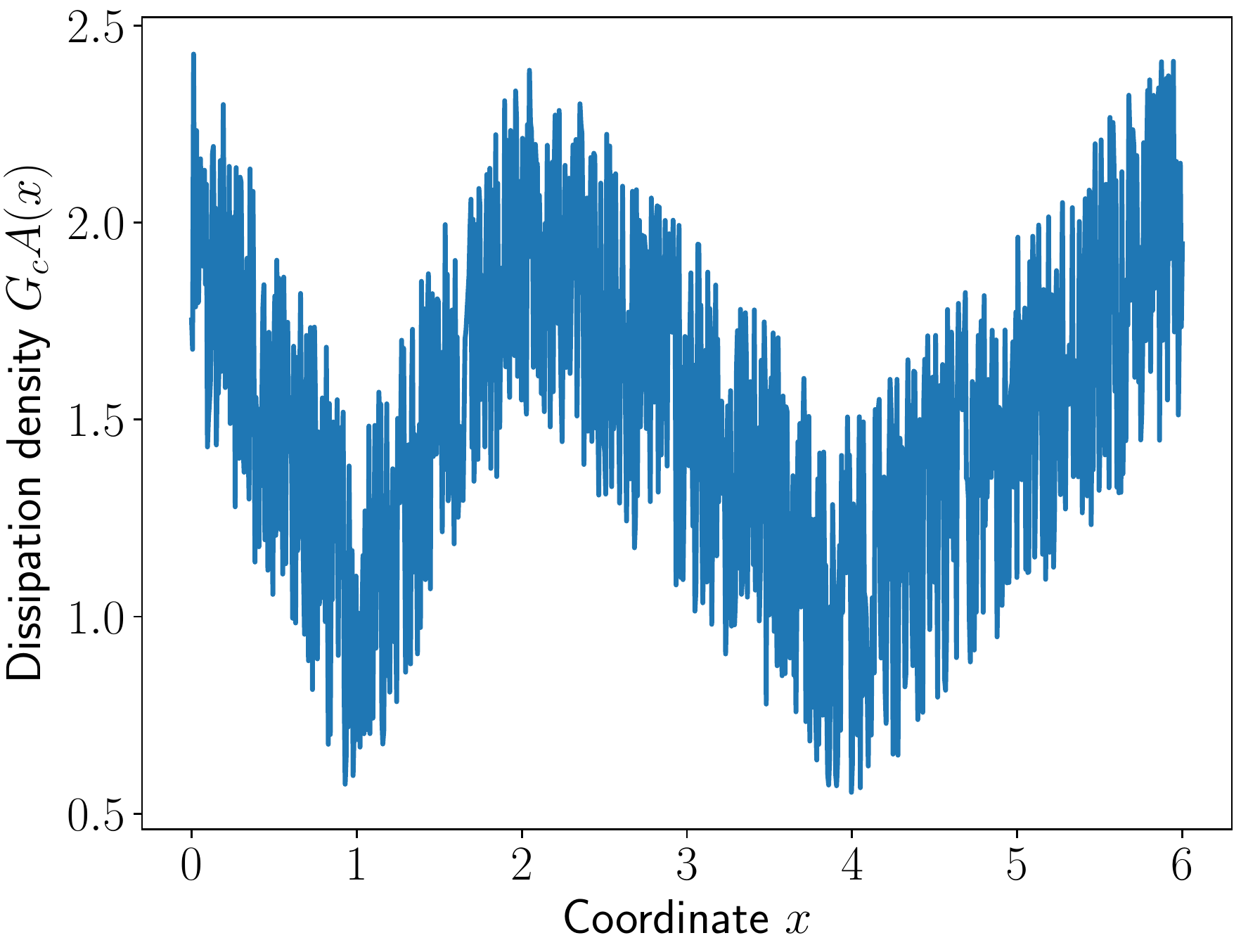}}
 \\
 \subfloat[]{\includegraphics[width=0.45\textwidth]{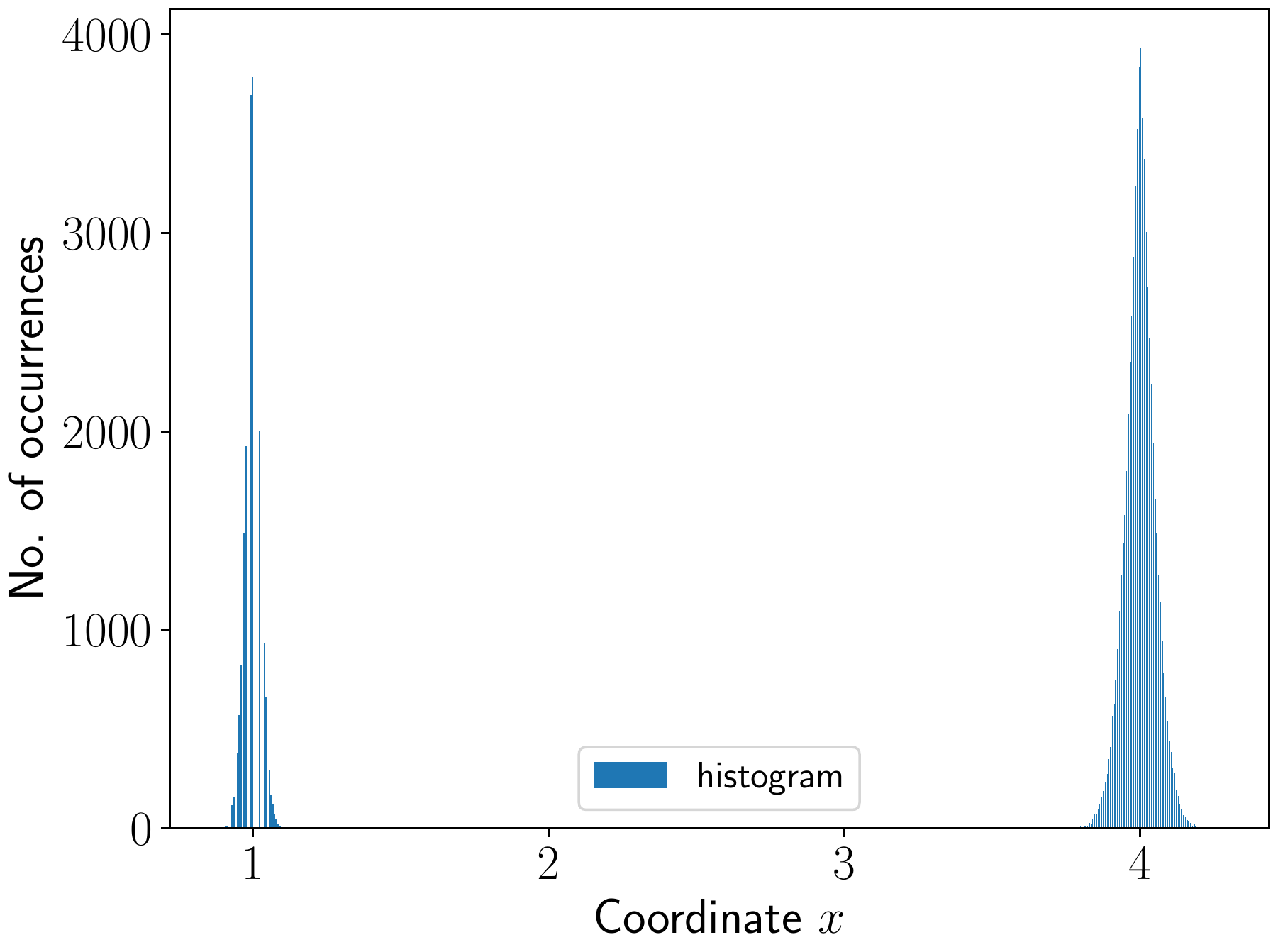}\label{fig:EA_min_hist}}
 \subfloat[]{\includegraphics[width=0.45\textwidth]{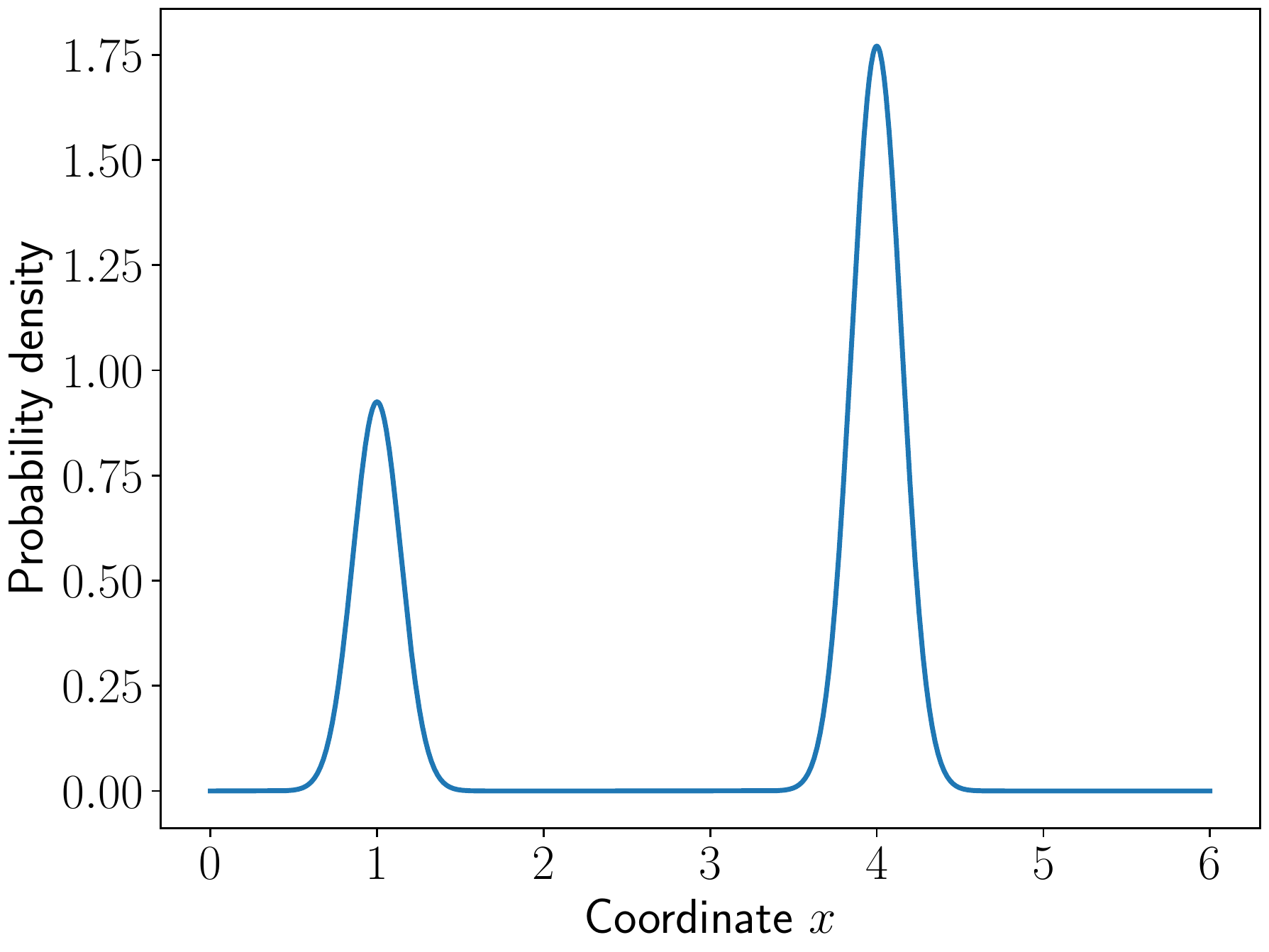}}
 \caption{Distribution of (a) dissipation density and (b) discretised dissipation density perturbed by 
uniform white noise, (c) histogram of crack points with (d) corresponding estimation of the probability density function for $\eta=0.01$. The probability of localisation is 1/3 for the neighbourhood of $x=1$, and 2/3 for the neighbourhood of $x=4$.}
 \label{fig:loc1}
\end{figure}

Considering a variable dissipation density, see Figure~\ref{fig:loc1}(a), such as
\begin{align}
\label{eq:yield1}
G_c A(x) = 
\begin{cases}
1+|x-1|&\text{for }x<2\\
1+\frac{|x-4|}{2}&\text{for }x\geq 2\\
\end{cases},
\end{align}
the bar breaks either at $x=1$ or at $x=4$. However, the probability of failure for individual points remains unknown. In order to obtain more information about these probabilities, we perturb the problem by random noise, which may reflect reality when e.g.\ the local behaviour is influenced by small heterogeneities.

Here the cross-sectional area $A(x)$ is perturbed with a white noise RV $\rv{q}$ with a uniform distribution supported on the interval $[-1/2,1/2]$.
Technically, at the discrete level at the grid points $x_i$ the dissipation is set
to $G_cA(x_i) + \eta\cdot\rv{q}_i$, where the uniform RV $\rv{q}_i$ is independent of 
the RVs $\rv{q}_j$ at other grid points $x_j$, $j \ne i$, and $\eta>0$
controls the magnitude of the perturbation.  In particular, $1001$ points were used
in the discretisation of the computational interval $[0,6]$.

The point of failure is simply computed as the minimum of the dissipation density.
The histogram of failure points and the corresponding probability density (estimated through kernel density estimation) are shown in Fig~\ref{fig:loc1}(c,d).
The failures occur in the neighbourhood of the points $1$ and $4$.
However, the dissipation density has a different shape around those points, which results in different probability distributions. 
In particular, the probability of a failure around the point $1$ is $1/3$, while it is $2/3$ around the point $4$, independently of the perturbation parameter $\eta$. These results suggest the existence of a unique relaxed minimizer as the distribution
\begin{equation} \label{eq:minimiz}
  \mv_{\rv{x}} = \frac{1}{3}\, \updelta_1 + \frac{2}{3}\, \updelta_4 .
\end{equation}

Note that the spatial discretisation has to be fine enough to obtain accurate numerical approximations of the probabilities for small perturbation parameters $\eta$.

\begin{figure}[htb]
 \centering
 \subfloat[]{\includegraphics[width=0.45\textwidth]{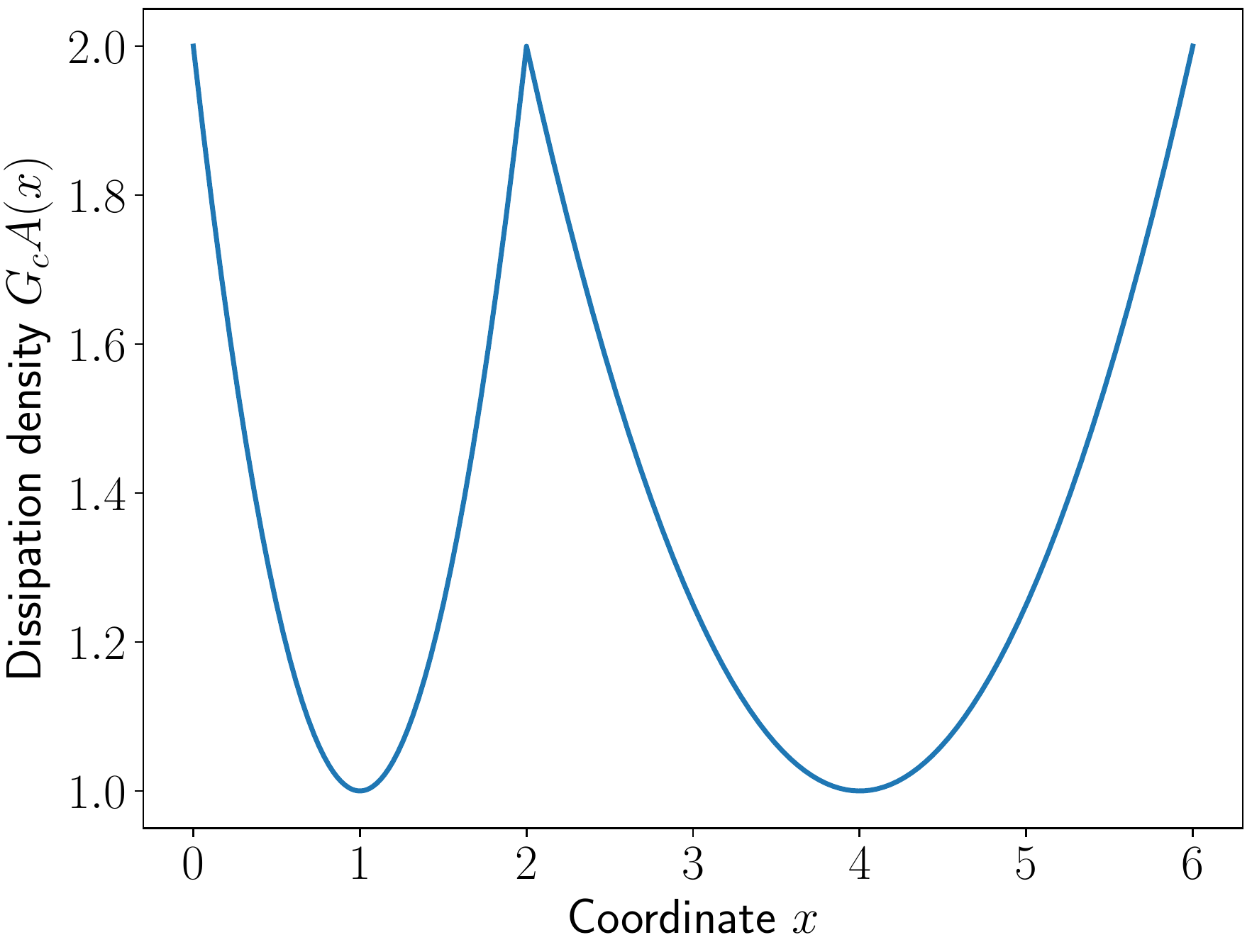}}
 \subfloat[]{\includegraphics[width=0.45\textwidth]{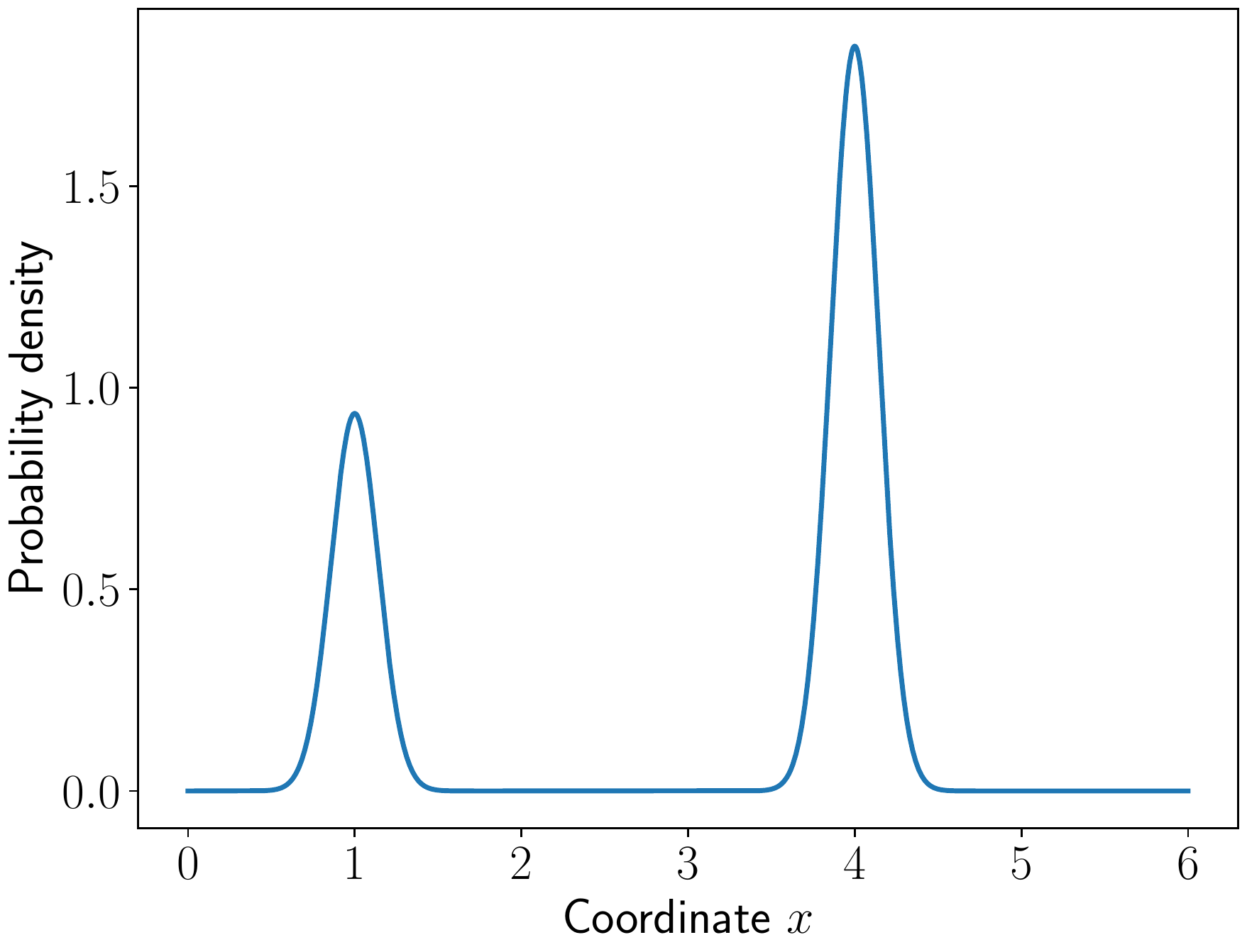}}
 \caption{Distribution of (a) dissipation density and (b) probability density of the crack points for $\eta=0.01$. The probability of localisation remains 1/3 for the neighbourhood of $x=1$, and 2/3 for the neighbourhood of $x=4$.}
 \label{fig:loc3}
\end{figure}

In the first example the shape of the dissipation density curve was like two
V-shaped notches. The next example has the dissipation density curve like two U-shaped notches, 
depicted in {Figure~\ref{fig:loc3}}, and defined as
\begin{align}
\label{eq:yield3}
G_cA(x) = 
\begin{cases}
1+(x-1)^2&\text{for }x<2\\
1+\frac{1}{4}(x-4)^2&\text{for }x\geq 2\\
\end{cases}.
\end{align}
Similarly to the previous example, the different shape of the fracture dissipation density curve gives rise to a different distribution of the crack points. However, the probabilities that the crack is located either at $x=1$ or $x=4$ remain 1/3 and 2/3 respectively, and are again independent of the perturbation parameter $\eta$. Once again this suggests a unique relaxed minimizer as in  \eqref{eq:minimiz}.

\begin{figure}[htb]
 \centering
 \subfloat[]{\includegraphics[width=0.45\textwidth]{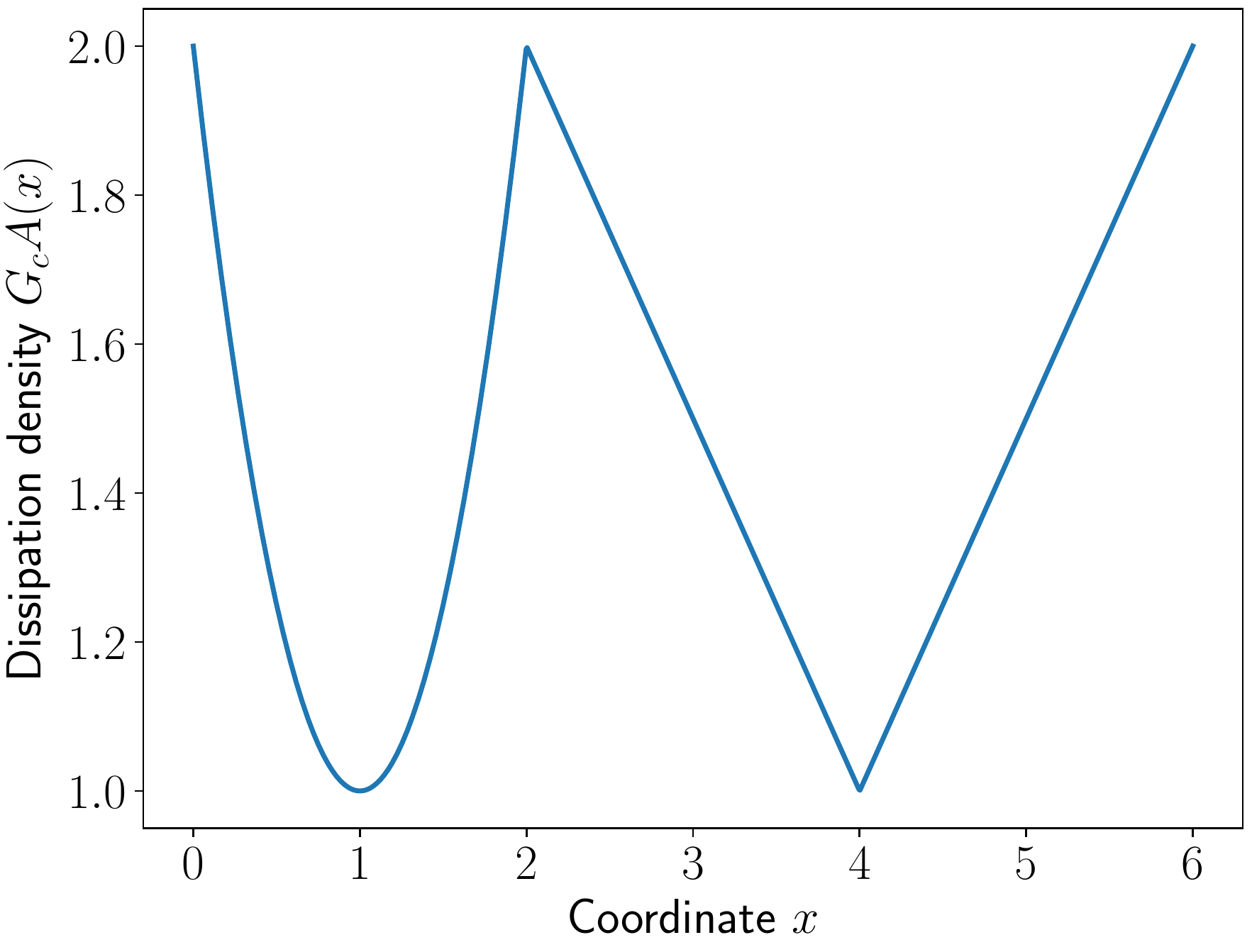}}
 \subfloat[]{\includegraphics[width=0.45\textwidth]{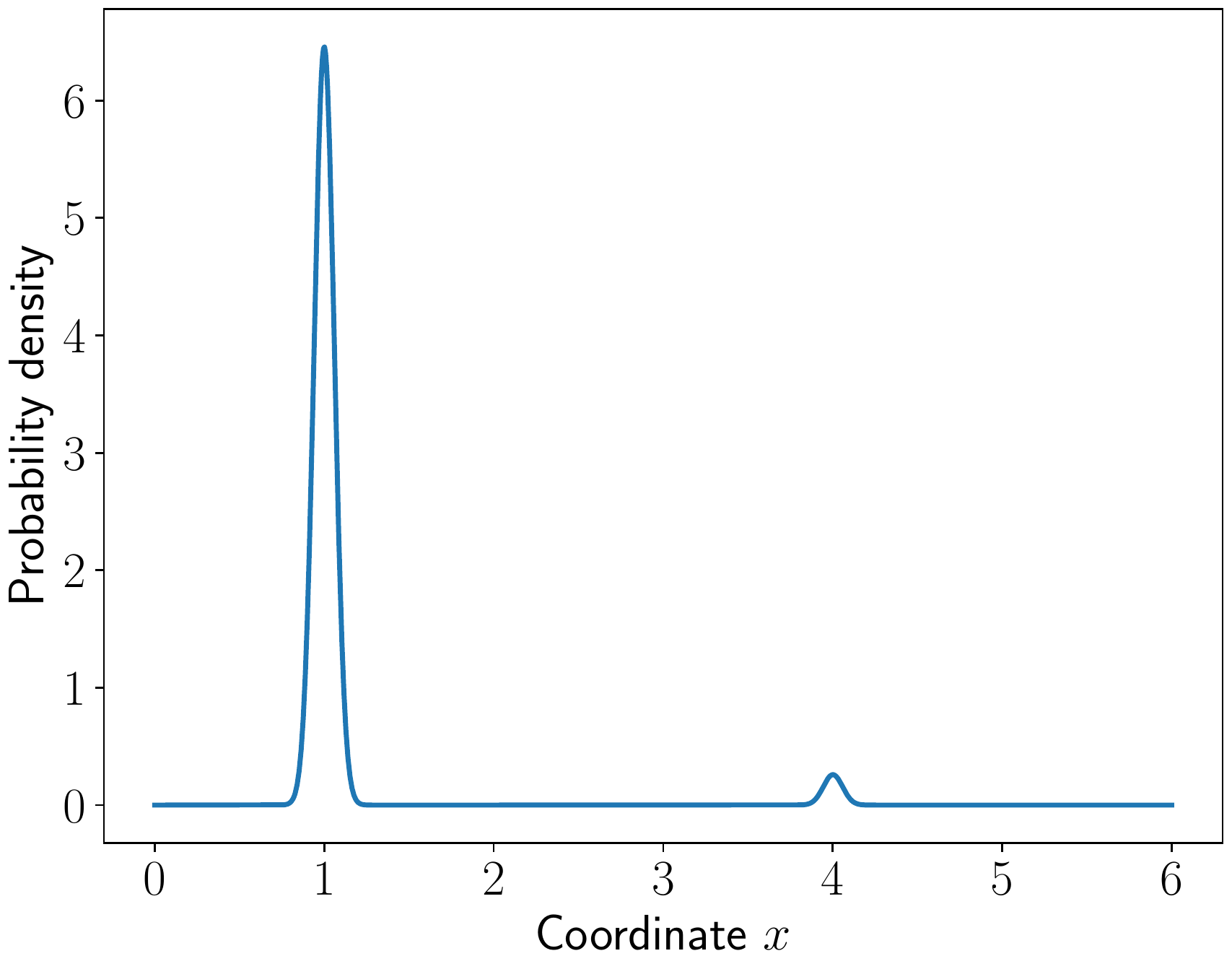}}
\caption{Distribution of (a) dissipation density, and (b) probability density of the crack points for $\eta=0.01$. The probability of localisation is attracted to the neighbourhood of $x=1$ with limiting probability equal to $1$.}
 \label{fig:loc2}
\end{figure}

The last example in this section has the dissipation density curve, shown in Fig.~\ref{fig:loc2}, in form of a U-shaped and V-shaped notch as
\begin{align}
\label{eq:yield2}
G_cA(x) = 
\begin{cases}
1+(x-1)^2&\text{for }x<2\\
1+\frac{|x-4|}{2}&\text{for }x\geq 2\\
\end{cases}.
\end{align}
Contrary to the previous two examples, the localisation of the failure point is attracted to the point $x=1$ for decreasing values of $\eta$, attaining the probability $1$ in the limit, so that there seems to be a unique relaxed minimizer with distribution  $\mv_{\rv{x}} = \updelta_1$.

\subsection{Fracture of a 1D bar in the phase-field setting}
\label{secPF_1D}
The example of the previous Subsection~\ref{sec:1D_griffith} is now analyzed in
the framework of phase field regularisation. The energetic functional then reads
\begin{equation*}
    \En(u,\alpha) = \frac{1}{2}\int_{0}^L (1-\alpha(x))^2  Y A(x) (u'(x))^2 \ \mathrm{d}x + \frac{1}{2} \int_{0}^L G_c A(x) \left( \frac{\alpha^2(x)}{\ell} + \ell (\alpha'(x))^2  \right) \D{x}.
\end{equation*}
Note that for this monotonic tension setup the irreversibility constraint for $\alpha$ does not need to be enforced as it is automatically satisfied. In the numerical tests we set $Y=10^4,L=6$ and use $G_c A(x)$ as given in \eqref{eq:yield1}, i.e. the double V-notch case. A uniform finite element mesh in $[0,L]$ is employed. The loading step is chosen as $\Updelta u = 0.1$ and we consider $10$ loading steps in total. At every fixed step, the staggered iterative solution process is carried out until $|\boldsymbol{\alpha}^{(k+1)} - \boldsymbol{\alpha}^{(k)}|< 10^{-4}$, where $\boldsymbol{\alpha}^{(k)}$ represents the nodal phase field values $| \cdot |$ is the Euclidean norm.

\begin{figure}[t!]
\begin{minipage}{0.49\textwidth}
 \subfloat[]{
\begin{tikzpicture}[scale = 0.85]
\begin{axis}[legend pos = north east, xlabel = Coordinate $x$, ylabel= Phase field variable $\alpha$]
\addplot[color = blue,thick] table [x=x, y=alpha, col sep=comma] {images/1D_PF_Sample_crack1.csv};
\end{axis}
\end{tikzpicture}
}
\end{minipage}
\begin{minipage}{0.49\textwidth}
 \subfloat[]{
\begin{tikzpicture}[scale = 0.85]
\begin{axis}[legend pos = north east, xlabel = Coordinate $x$, ylabel= Phase field variable $\alpha$]
\addplot[color = blue,thick] table [x=x, y=alpha, col sep=comma] {images/1D_PF_Sample_crack2.csv};
\end{axis}
\end{tikzpicture}
}
\end{minipage}
\caption{Different crack types for the 1D phase field example.}
    \label{fig:1D_PF_cracks}
\end{figure}
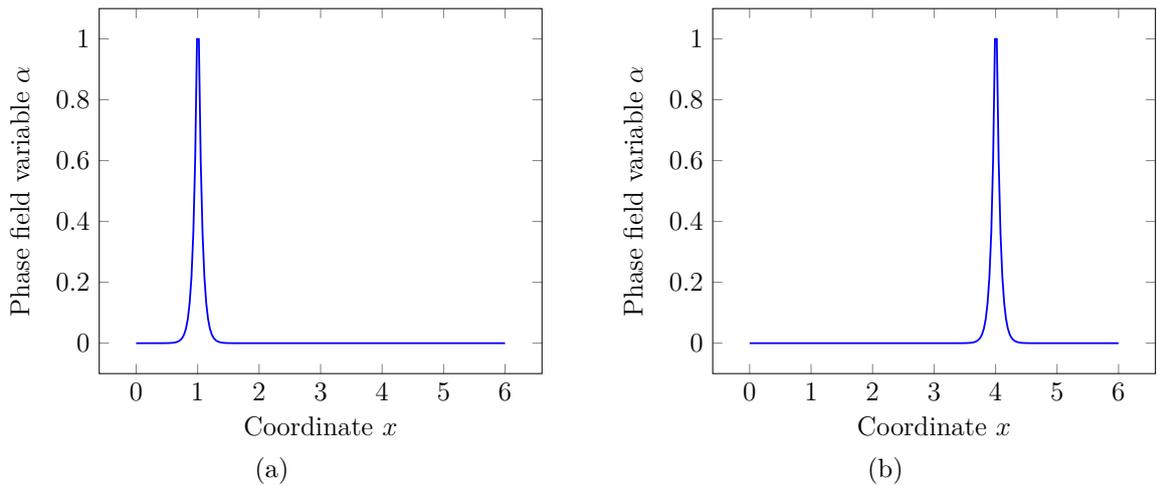

Unlike in the previous example, here we look for solutions which correspond to local minima (or even stationary points) of the energy, as our numerical procedure solves the Euler-Lagrange equations. For a thorough discussion of the differences between predictions of the two approaches in the context of fracture, see \cite{Negri2008}. The perturbation of $G_c A(x)$ is generated in the same way as in Section~\ref{sec:1D_griffith}: at the grid nodes $x_i$ the dissipation is set to $G_cA(x_i) + \eta\cdot\rv{q}_i$ with $\rv{q}_i$ denoting pairwise independent and uniformly distributed random numbers in the interval $[-1/2,1/2]$. Then, a piecewise linear interpolation is used to obtain the perturbed $G_c A(x)$. Since computations are more expensive here than in the previous two examples, we choose a smaller sample size, but monitor the sampling accuracy with the $95\%$ confidence interval $\Delta_{0.95} = 1.96 \, \sqrt{p_{i} (1-p_{i})/M}$, $i=1,2$, see \cite{Tyagi2018}. Note that $p_1$ and $p_2$ represent the probabilities of cracks occuring at $x=1$ and $x=4$, respectively. Figure \ref{fig:1D_PF_cracks} depicts the two cracks appearing in this scenario.

\begin{table}[h!]
\centering
\begin{tabular}{|c|c|c|c|c|c|}
\hline
$\eta$ & $\ell$ & $\#$ grid points & $p_1$ & $p_2$ \\
\hline 
\hline
$1$ &  $0.1 L$ & $500$ & $0$ & $1$\\
\hline
$1$ &  $0.01 L$ & $1000$ & $0.24$ & $0.76$\\
\hline
$1$ &  $0.001 L$ & $2000$ & $0.33$ & $0.67$ \\
\hline 
\hline
$0.5$ &  $0.01 L$ & $1000$ & $0.11$ & $0.89$ \\
\hline
$0.5$ &  $0.001 L$ & $2000$ & $0.32$ & $0.68$ \\
\hline
$0.5$ &  $0.0001 L$ & $5000$ & $0.34$ & $0.66$  \\
\hline
\hline 
$0.1$ &  $0.01 L$ & $1000$ & $0$ & $1$ \\
\hline
$0.1$ &  $0.001 L$ & $2000$ & $0.24$ & $0.76$ \\
\hline
$0.1$ &  $0.0001 L$ & $5000$ & $0.34$ & $0.66$ \\
\hline 
\end{tabular}
\caption{Crack probabilities obtained with the phase field approach and Monte Carlo sampling for varying perturbation size, regularization length and grid interval size. In all cases, the sample size is $M=2500$ and for the sampling error there holds $\Delta_{0.95} \leq 0.02$.}
\label{tab:PF_1d}
\end{table}

Table~\ref{tab:PF_1d} shows the computed probabilities for different perturbation magnitudes. For all settings a sample size of $M=2500$ has been employed for which the $95 \%$ confidence interval is smaller than $0.02$. The results reveal that the regularization length $\ell$ and the mesh interval size have to be chosen small enough to capture the stochastic perturbations. For instance, when a relatively large perturbation ($\eta=1$) is resolved with $\ell = 0.01 L$ and $1000$ grid points (the same number of points as in Section~\ref{sec:1D_griffith}), we observe a bias in the computed probabilities. Only after refining $\ell$ and the mesh size we obtain $p_1=1/3$ and $p_2=2/3$, as expected. Note that when $\ell$ is decreased, a finer mesh is needed to properly resolve the regularized cracks. The same observations can be made for smaller perturbation amplitudes, where the mesh interval size and $\ell$ have to be successively refined to recover $p_1=1/3$ and $p_2=2/3$. If the resolution is much too coarse, only cracks at $x=4$ are observed, since the perturbation cannot be resolved by the phase field finite element method. Indeed, a crack at $x=4$ is also obtained for our specific implementation in this example, if all coefficients are deterministic. These results suggest that there exists again a unique relaxed minimizer as in  \eqref{eq:minimiz}.

\section{Stochastic phase-field modeling of brittle fracture}
\label{secProb_gen}

As mentioned earlier, the minimisation of non-convex energy functionals such as
\eqref{RegVAF} may produce local minima.  Especially when the energy levels of these competing minima are close, little random perturbations in the physical system and/or in the numerical setup may lead to different solutions as has been demonstrated in the examples shown. This is further complicated by the fact
that the necessary condition \eqref{Weak} of the vanishing first variation has all stationary points of the functional as solutions.  The idea is
therefore to relax the minimisation problem  in such a way as to hopefully capture all these possibilities. This will be done explicitly be assuming some random
perturbations in the energy functional \eqref{RegVAF} by allowing some quantity specifying the functional to be a RV, or more generally a random field (RF).

In this way, the energy functional \eqref{RegVAF} becomes a RV, and one has to specify what it means to minimise it.  Here, guided by previous work on 
variational stochastic extensions for elasticity
\cite{hgmCbu99,babuska2004stochastic,matthies2005stochastic}
as well as for plasticity \cite{HgmRos08a,bvrHgm11,BVRhgm2015} in a convex analysis framework, a new stochastic energy functional is defined as the expected value of the randomised deterministic energy functional \eqref{RegVAF}. From here on things can proceed in a theoretically
analogous manner to the deterministic formulation in section~\ref{SSecBrittle}. 

In this section, we first introduce some necessary concepts on RVs and probability. Then, we formulate the proposed stochastic phase-field model and computational framework for brittle fracture, which we finally illustrate with numerical results.

\subsection{Preliminaries}
\label{sec:random-variables-and-probability-distributions}

As follows, we briefly introduce the concepts and the notation needed for the formulation of the proposed stochastic phase-field model in the following section.

\subsubsection{Random variables and random fields }

We start with the familiar concept of scalar or real valued random variables (RVs) and their expectation.
Formally (see e.g.\ \cite{Sullivan2015}), such RVs can be represented as measurable real
valued functions on a probability space $(\Theta,\F{S},\P)$, where $\Theta$
is the set of all possible samples or realisations, and $\F{S}$
is a $\sigma$-algebra of measurable subsets of $\Theta$ --- the so-called events --- 
on which the probability measure $\P$ is defined.

What is important here for our purposes is that such random 
variables (RVs) form a vector space on which the expectation is defined as a
positive linear functional via the integral w.r.t.\ the probability measure, which for such a
RV $\tns{a}:\Theta\to\B{R}$ is given in the usual way
by $\E{\tns{a}} := \int_\Theta \tns{a}(\theta) \,\P(\di \theta) \in \B{R}$.
In case the RV $\tns{a}$ has a density $f_{\tns{a}}$ on its range $\B{R}$, one also has
the familiar relation $\E{\tns{a}} = \int_{\B{R}} \eta f_{\tns{a}}(\eta)\, \di\eta$.
Note that the probability of some event $\C{E}\in\F{S}$ can be stated
as an expectation $\P(\C{E}) = \E{\ch{\C{E}}}$, where the indicator
or characteristic function of $\C{E}$ is a RV which takes
the value one ($\ch{\C{E}}(\theta)=1$) if $\theta\in\C{E}$, and vanishes otherwise.
The mean of a RV $\tns{a}$ is often denoted by $\bar{\tns{a}}:=\E{\tns{a}}\in\B{R}$,
and the mean-free fluctuating random part by $\tilde{\tns{a}}:= \tns{a} - \bar{\tns{a}}$.
The real RVs form not only a vector space, but an algebra, as one may
define a product simply in the usual way by point-wise multiplication, and in this way
obtain an inner product $\bkt{\tns{a}}{\tns{b}}_{\C{S}} := \E{\tns{a}\tns{b}}$.
The usual Hilbert Lebesgue space of RVs with finite variance, 
which will be used later, is denoted by
\begin{equation}  \label{eq:S-def}
\C{S} := \Lp_2(\Theta;\B{R}) = \{ \tns{a}:\Theta\to\B{R}:\; \nd{\tns{a}}_{\C{S}}^2 := 
\bkt{\tns{a}}{\tns{a}}_{\C{S}} = \E{\tns{a}^2} < \infty \}.
\end{equation}
For RVs $\tns{a}, \tns{b}\in\C{S}$ the covariance is given by $\cov(\tns{a},\tns{b}):=
\bkt{\tilde{\tns{a}}}{\tilde{\tns{b}}}_{\C{S}} = \E{\tilde{\tns{a}}\tilde{\tns{b}}}$, and
the variance and standard deviation by $\MR{var}(\tns{a}):= \cov(\tns{a},\tns{a})=
\nd{\tilde{\tns{a}}}_{\C{S}}^2 = \bkt{\tilde{\tns{a}}}{\tilde{\tns{a}}}_{\C{S}}=
\E{\tilde{\tns{a}}^2}$ and $\MR{std}(\tns{a}):= \sqrt{\MR{var}(\tns{a})}$.  In case 
$\cov(\tns{a},\tns{b})=0$, the RVs $\tns{a}$ and $\tns{b}$ are called {\em uncorrelated}.
%}

Two RVs $\rv{a}_1$ and $\rv{a}_2$, possibly defined on different
probability spaces $\Theta_1$ and $\Theta_2$, are considered equivalent if 
$\E{\varphi(\rv{a}_1)} = \E{\varphi(\rv{a}_2)}$
for all functions $\varphi$ where that expression makes sense.  This means in particular
that equivalent RVs have the same distribution and moments. Similarly, two such
RVs are independent if
\[
  \E{\varphi(\rv{a}_1)\varphi(\rv{a}_2)} = \E{\varphi(\rv{a}_1)}\E{\varphi(\rv{a}_2)}
\]
for all functions $\varphi$ where that expression makes sense.  Note that independent
RVs are always uncorrelated, but the reverse implication may not hold.

The next task is to formalise RFs, e.g.\ RVs with values in the 
Hilbert Sobolev space $\sP$ in \eqref{eq:Pn}.  These will be possible candidates for 
minimisers of the stochastic variational formulation, and such RFs are also used to 
perturb the energy functional and are an input to the stochastic formulation and computation.
A RF $\tnb{a}(\theta,\vek{x}):\Theta\times\Dom\to\B{R}$ is a function of two arguments, 
namely $\theta\in\Theta$ as the stochastic variable and $\x\in\Dom$ as the spatial variable.
For  $\hat{\theta}\in\Theta$ fixed, $\tnb{a}(\hat{\theta}):=\tnb{a}(\hat{\theta},\cdot)$ 
is a deterministic phase field which we will want to be in $\sP$, whereas
on the other hand, for $\hat{\x}\in\Dom$  fixed, $\tnb{a}(\cdot,\hat{\x})$
is a real valued RV which we will want to be in $\C{S}=\Lp_2(\Theta;\B{R})$.

\subsubsection{Random phase fields}
As very general RVs with values in an infinite dimensional Hilbert space may pose 
some unexpected mathematical difficulties, we restrict ourselves to a somewhat
simple situation \cite{Besold2000,matthies2005stochastic} which is general
enough to display the idea.  From the Hilbert space $\C{S}$ of RVs in \eqref{eq:S-def} 
and the Hilbert space $\sP=\Hp^1(\Dom;\B{R})$ in \eqref{eq:Pn} of deterministic phase fields
with the usual Sobolev inner product $\bkt{\cdot}{\cdot}_{\sP}$, 
we form a new Hilbert space of random phase fields (RFs) as the Hilbert tensor
product of possible solutions to the stochastic variational problem
(cf.\ \cite{Besold2000,matthies2005stochastic,BVRhgm2015})
\begin{equation}  \label{eq:RFs-pha}
   \rsP := \C{S}\otimes\sP \quad\text{ with usual inner product }\quad 
   \bkd{\tnb{a}}{\tnb{b}}_{\rsP} := \bkt{\tns{a}}{\tns{b}}_{\C{S}} \bkt{\alpha}{\beta}_{\sP}
\end{equation}
for elementary tensors $\tnb{a}=\tns{a}\otimes\alpha, \tnb{b}=\tns{b}\otimes\beta
\in \rsP := \C{S}\otimes\sP$, where for example in $\tnb{a}=\tns{a}\otimes\alpha$ 
--- ($\tnb{a}(\theta,\x)=\tns{a}(\theta)\alpha(\x)$) --- the
factor $\tns{a}\in\C{S}$ is a RV and the factor $\alpha\in\sP$ is a deterministic
phase field.  As the whole space $\rsP$ is composed of sums and convergent
series $\sum_j \tns{a}_j\otimes\alpha_j$ of such elementary tensors, 
the inner product in \eqref{eq:RFs-pha} is extended by linearity to the whole space.
Note that \eqref{eq:RFs-pha} implies that the induced $\rsP$-norm is a cross norm:
$\nt{\tnb{a}}_{\rsP} = \nd{\tns{a}}_{\C{S}} \nd{\alpha}_{\sP}$.
If the $\sP$-inner product and -norm are extended to $\rsP$ in the obvious fashion,
namely for elementary tensors like above
$\bkt{\tnb{a}(\theta,\cdot)}{\tnb{b}(\theta,\cdot)}_{\sP}:= 
    \tns{a}(\theta)\tns{b}(\theta) \bkt{\alpha}{\beta}_{\sP}$ and
    $\nd{\tnb{a}(\theta,\cdot)}_{\sP} = \ns{\tns{a}(\theta)} \nd{\alpha}_{\sP}$,
then $\bkd{\tnb{a}}{\tnb{b}}_{\rsP} = \E{\bkt{\tnb{a}}{\tnb{b}}_{\sP}}$ and
$\nt{\tnb{a}}_{\rsP} = \E{\nd{\tnb{a}}_{\sP}}$, i.e.\ the quantities on the stochastic
space $\rsP$ are just the expectations of the ones on the base space $\sP$,
a pattern which will be repeated several times.

Similarly, the expectation as a linear map
to the basis space $\sP$ of deterministic fields is defined on $\rsP$ by 
$\bar{\tnb{a}}=\E{\tnb{a}} := \E{\tns{a}} \alpha \in \sP$  on elementary tensors
$\tnb{a}=\tns{a}\otimes\alpha\in\rsP$, and extended to all of $\rsP$
by linearity.  It is well known \cite{Sullivan2015} that the tensor product
space $\rsP$ in \eqref{eq:RFs-pha} is isomorphic with the Hilbert space of RFs
$\tnb{a}:\Theta\to\sP$ where the $\sP$-norm has finite variance
\begin{multline}  \label{eq:RFs-pha-L2}
  \rsP = \C{S}\otimes\sP = \Lp_2(\Theta;\B{R})\otimes\Hp^1(\Dom;\B{R})\cong 
  \Lp_2(\Theta;\sP) = \Lp_2(\Theta;\Hp^1(\Dom;\B{R})) \\  
  = \{ \tnb{a} :\;  \nt{\tnb{a}}_{\rsP}^2 := \bkd{\tnb{a}}{\tnb{a}}_{\rsP}
  = \E{\bkt{\tnb{a}}{\tnb{a}}_{\sP}} = \E{\nd{\tnb{a}}_{\sP}^2}  < \infty \}.
\end{multline}

\subsubsection{Quantities of interest}
A quantity of interest (QoI) in the deterministic setting is typically some function
$\Upsilon(\Vu,\alpha)$ of the solution $(\Vu,\alpha)\in\sU_n\times\sP_n$.  Now if
in the stochastic formulation the deterministic fields $(\Vu,\alpha)$ are replaced
by RFs $(\uu,\tnb{a})$, and inserted into the deterministic QoI $\Upsilon(\uu,\tnb{a})$,
this becomes a RV.  One is then typically interested in a new QoI like
$\vek{\Upsilon}(\uu,\tnb{a}) := \E{\Upsilon(\uu,\tnb{a})}$, where $\Upsilon$ is an appropriate function.  Some examples are the mean 
$\bar{\tnb{a}}=\E{\tnb{a}}$ with $\Upsilon(\uu,\tnb{a}) = \tnb{a}$, 
or the $p$-th central moment $\E{(\tnb{a} - \E{\tnb{a}})^p}$ with
$\Upsilon(\uu,\tnb{a})= (\tnb{a} - \bar{\tnb{a}})^p$. As already mentioned, in general a pattern is that the probabilistic QoI is 
the expected value of the deterministic QoI.

\subsubsection{Separated representation}
The relations \eqref{eq:RFs-pha} and \eqref{eq:RFs-pha-L2} give also a practical way
of approximating RFs in a {\em separated representation} as linear combinations
$\tnb{a}(\theta,\vek{x}) \approx \sum_j \tns{a}_j(\theta) \alpha_j(\vek{x})$
of deterministic fields $\alpha_j\in\sP$ with RVs $\tns{a}_j\in\C{S}$ as coefficients.
Such separated expansions are typical for parametric maps $\tnb{a}:\Theta\to\sP$ 
from a set $\Theta$ into a Hilbert space $\sP$, which
can be analysed very generally in terms of linear operators \cite{hgmRO-1-2018-p}.
This kind of analysis was started in probability theory
in \cite{Karhunen1947,loeve1977probability,loeve1978probability} in
terms of the so-called Karhunen-Lo\`eve expansion \cite{Sullivan2015}.
It begins by defining for a RF $\tnb{a}$ a bilinear form for any 
$(\beta_1,\beta_2)\in\sP\times\sP$ by
\[
     \E{\ip{\tilde{\tnb{a}}}{\beta_1}\ip{\tilde{\tnb{a}}}{\beta_2}}
     = \ip{C_{\tnb{a}}\beta_1}{\beta_2} = \iint_{\Dom\times\Dom} \beta_1(\vek{x})\,
     \cov_{\tnb{a}}(\vek{x},\vek{y})\, \beta_2(\vek{y})\, \di\vek{x}\di\vek{y} ,
\]
which in turn defines the self-adjoint positive semi-definite
{\em covariance operator} $C_{\tnb{a}}:\sP\to\sP^*$ and the symmetric 
positive semi-definite correlation function $\cov_{\tnb{a}}:\Dom\times\Dom\to\B{R}$.  
This shows that $C_{\tnb{a}}$ can be represented as an integral operator, 
and its eigenvalues equal those of the integral operator with kernel equal 
to $\cov_{\tnb{a}}$.  As $\cov_{\tnb{a}}(\vek{x},\vek{y}) = 
\E{\tilde{\tnb{a}}(\cdot,\vek{x}) \tilde{\tnb{a}}(\cdot,\vek{y})} = 
\bkt{\tilde{\tnb{a}}(\cdot,\vek{x})}{\tilde{\tnb{a}}(\cdot,\vek{y})}_{\C{S}}=
\cov(\tnb{a}(\cdot,\vek{x}),\tnb{a}(\cdot,\vek{y}))$,
it is easily seen that the local variance and standard variation at $\x\in\Dom$ are 
$\MR{var}_{\tnb{a}}(\x)=\cov_{\tnb{a}}(\x,\x)$ and 
$\MR{std}_{\tnb{a}}(\x)=\sqrt{\MR{var}_{\tnb{a}}(\x)}$, and 
the total variance of the RF $\tnb{a}$ is defined as
\begin{equation}  \label{eq:tot-var-pha}
\B{V}(\tnb{a}) := \E{\ip{\tilde{\tnb{a}}}{\tilde{\tnb{a}}}} = 
\int_{\Dom} \MR{var}_{\tnb{a}}(\vek{x}) \, \di\vek{x} = 
\int_{\Dom} \cov_{\tnb{a}}(\vek{x},\vek{x}) \, \di\vek{x} = \MR{tr}\, C_{\tnb{a}} <\infty.
\end{equation}
This means that $C_{\tnb{a}}$ is a compact operator, where the sum of eigenvalues 
(the trace) is finite, i.e.\ $C_{\tnb{a}}$ is a trace-class or nuclear operator, and
thus the RF $\tnb{a}\in\rsP$ represents a proper RV, i.e.\ a 
measurable map $\tnb{a}:\Theta\to\sP$ \cite{Sullivan2015}. The eigenvalue equation for $   C_{\tnb{a}}(\varphi_j)(\vek{x}) = 
\int_{\Dom} \cov_{\tnb{a}}(\vek{x},\vek{y})\, \varphi_j(\vek{y})\, \di\vek{y} = 
\lambda_j \varphi_j(\vek{x})$ then leads to the celebrated Karhunen-Lo\`eve expansion
\cite{Karhunen1947,loeve1977probability,loeve1978probability,Sullivan2015},
a separated expansion in terms of orthogonal eigenfunctions of $C_{\tnb{a}}$:
\begin{equation}   \label{eq:KLE}
   \tnb{a}(\theta,\vek{x}) = \bar{\tnb{a}}(\vek{x}) + \sum_{j=1}^\infty \sqrt{\lambda_j}\,
   \zeta_j(\theta) \varphi_j(\vek{x}).
\end{equation}
This expansion can be shown to correspond to a singular value decomposition \cite{hgmRO-1-2018-p}.
The uncorrelated zero mean unit variance RVs $\zeta_j\in\C{S}$ are given by orthogonal projection
$\zeta_j(\theta) = \ip{\tilde{\tnb{a}}(\theta,\cdot)}{\varphi_j} = \int_{\Dom}
\tilde{\tnb{a}}(\theta,\vek{y}) \varphi_j(\vek{y}) \, \di\vek{y}$.  Arranging the
eigenvalues $\lambda_j$ of $C_{\tnb{a}}$ in a descending order, the truncation of 
the Karhunen-Lo\`eve series \eqref{eq:KLE} after $J$ terms gives
the \emph{best} $J$-term approximation to $\tnb{a}$, and is often used in the
generation resp.\ sampling of RFs.

\subsubsection{Random displacement fields}
A completely analogous construction is carried out for the displacements, so that one
arrives at a Hilbert space of stochastic displacement variations $\rsU = \C{S}\otimes\sU$ as
the probabilistic analogue of the deterministic Hilbert space $\sU$ in \eqref{eq:U-def}.
In particular, as the deterministic space $\sU$ is a space of $\sR^d$-valued fields 
%--- with $\C{F}$ (typically $\C{F}=\sR^d$) the vector space for the displacements ---
the space $\rsU$ will be a Hilbert space with $\sR^d$-valued RFs. The inner
product on the deterministic space $\sU$ can be taken as $\bkt{\vek{u}}{\vek{v}}_{\sU} = 
\int_{\Dom} \MR{tr}\,(\nabla\vek{u}(\vek{x})^{\trpos}\cdot \nabla\vek{v}(\vek{x}))\,\di \vek{x}$,
and on the stochastic space $\rsU$ it is in analogy to \eqref{eq:RFs-pha} defined 
for elementary tensors
$\tnb{u}=\tns{u}\otimes\vek{u}, \tnb{v}=\tns{v}\otimes\vek{v}\in \rsU = \C{S}\otimes\sU$
as $\bkd{\tnb{u}}{\tnb{v}}_{\rsU}:=\bkt{\tns{u}}{\tns{v}}_{\C{S}}\bkt{\vek{u}}{\vek{v}}_{\sU}
= \E{\bkt{\tnb{u}}{\tnb{v}}_{\sU}}$, and extended by linearity.  Once again one has a congruence
like \eqref{eq:RFs-pha-L2}:
\begin{align}  \label{eq:RFs-dsp-L2}
  \rsU &= \C{S}\otimes\sU = \Lp_2(\Theta;\B{R})\otimes\Hp^1_\Gamma(\Dom;\sR^d)\\ 
  \nonumber
  & \cong \Lp_2(\Theta;\B{R})\otimes\Hp^1_\Gamma(\Dom;\B{R})\otimes\C{F} 
  \cong \Lp_2(\Theta;\sU) = \Lp_2(\Theta;\Hp^1_\Gamma(\Dom;\C{F})) \\
  \nonumber
   &= \{ \tnb{v} :\;  \nt{\tnb{v}}_{\rsU}^2 :=
  \bkd{\tnb{v}}{\tnb{v}}_{\rsU} = \E{\bkt{\tnb{v}}{\tnb{v}}_{\sU}}
  = \E{\nd{\tnb{v}}_{\sU}^2} < \infty \}.
\end{align}
All the other following constructions can be carried out in a completely analogous fashion
to the ones for the space of random phase fields $\rsP$ and need not be repeated here.

\subsubsection{Stochastic constraints}
It remains to define the stochastic analogues of the affine space $\sU_n$ in \eqref{eq:Un}
and the convex set $\sP_n$ in \eqref{eq:Pn}.  These will be used in the stochastic
variational problems to be considered in 
Section~\ref{sec:stochastic-formulations-of-variational-problems}.
The stochastic affine space $\rsU_n$ is defined as
\begin{multline}   \label{eq:rUn}
   \rsU_n := \{ \tnb{u}\in\Lp_2(\Theta;\Hp^1(\Dom;\sR^d)):    \\  
   [ \tnb{u}(\theta,\vek{x}) = 0  \text{ on } \Gamma_{\MR{Dir},0}\; \text{ and } \;
   \tnb{u}(\theta,\vek{x}) = \bar{\vek{u}}_n \text{ on } \Gamma_{\MR{Dir},1} ]
   \; \P-\text{a.s.} \} .
\end{multline}
The stochastic analogue of the convex subset $\sP_n \subset \sP$ is {again a convex subset of $\rsP$}, and is defined as
\begin{equation} \label{eq:rPn}
  \rsP_n := \{\tnb{a}\in\rsP:\; [\tnb{a}(\theta,\cdot) \ge \tnb{a}_{n-1}(\theta,\cdot)]
         \; \P-\text{a.s.} \} \subset \rsP ,
\end{equation}
where $\tnb{a}_{n-1}\in\rsP$ is known from the previous time step. In stochastic optimisation, when a previously deterministic constraint 
condition becomes random, it is often a question which would be the best way to 
enforce this condition in the stochastic case \cite{Marti05}. Observe that it could
also be regarded as reasonable that instead of requiring e.g.\ in \eqref{eq:rUn} that
$\tnb{u}(\theta,\vek{x}) = 0  \text{ on } \Gamma_{\MR{Dir},0}$ almost surely (a.s.) in 
the measure $\P$ --- i.e.\ that the probability of this condition being violated vanishes 
--- one demands that only $\E{\tnb{u}(\cdot,\vek{x})}=0 \text{ on } \Gamma_{\MR{Dir},0}$.
This would have been a much laxer condition, but the formulation in \eqref{eq:rUn} seems
to be a natural generalisation of \eqref{eq:Un}, and demands the strict enforcement
of the boundary condition for every realisation.

\subsection{Stochastic formulation of the variational problem}\label{sec:stochastic-formulations-of-variational-problems}
The stochastic variational formulation to be presented here first needs to introduce
some probabilistic notion into the minimisation problem \eqref{argmin0}.  This will
be achieved by a small random perturbation in the definition of the functional $\En(\Vu,\alpha)$
by letting some variable or field $\Vq$ which appears in the definition of $\En$ become
a RV or RF $\tnb{q}$.  This way the functional now has become a
RV.  The second ingredient then will be to allow the solutions $(\Vu,\alpha)$ to be
RFs  $(\uu,\tnb{a})$ and define a new functional as the expectation of the
deterministic one.

This is then the setting for the stochastic variational problem. It is a new real valued
functional which has to be minimised over random fields.  The necessary conditions for
this will be derived similarly to \eqref{Weak} for the deterministic
minimisation problem \eqref{argmin0}. Thus, in this general case such as in the 1D example of Section \ref{secPF_1D}, we pursue the local minimization (or even the stationarity) problem and not the global minimization problem.

Let us start by choosing some quantity $\Vq$ which appears in the definition of 
$\En(\Vu,\alpha)=\En(\Vq;\Vu,\alpha)$
to be perturbed. This choice obviously depends on the particular form of the
energy functional $\En$. Then this quantity is replaced by a random one $\tnb{q}$, and the
new functional is now a RV $\En(\tnb{q};\Vu,\alpha)$, since $\tnb{q}$ is
random.  Following 
\cite{hgmCbu99,babuska2004stochastic,matthies2005stochastic,HgmRos08a,bvrHgm11,BVRhgm2015}
we choose the new energy functional to be the expected value of the deterministic one.
The last ingredient is to allow RFs $(\uu,\tnb{a})$ to be the solution.  With the preparations
in the previous Section~\ref{sec:random-variables-and-probability-distributions} and in particular with the
definition of the admissible sets of RFs $\rsU_n$ in \eqref{eq:rUn} and $\rsP_n$ in
\eqref{eq:rPn}, the new minimisation problem reads as follows:
\begin{align}  \label{eq:min_RV}
 (\uu,\tnb{a}) &= \argmin \{\Ens(\vv,\tnb{b}):\; \vv\in\rsU_n,\; \tnb{b}\in\rsP_n \}, \\ 
 \text{with } \Ens(\vv,\tnb{b}) &= \E{\En(\tnb{q};\vv,\tnb{b})}.  \label{eq:new-J}
\end{align}

As formally the situation is completely equal to the deterministic case, the derivation of
the necessary conditions is the same, giving in analogy to \eqref{Weak}
\begin{equation}
\left\{
\begin{tabular}{l}
$\ipd{\Gdiff_{\uu}\Ens(\uu,\tnb{a})}{\vv} := \E{\ip{\Gdiff_{\uu} \En(\tnb{q};\uu,\tnb{a})}{\vv}} = 0 
        \quad \forall \vv\in\rsU$,   \\[0.2cm]
$\ipd{\Gdiff_{\tnb{a}}\Ens(\uu,\tnb{a})}{\tnb{b}-\tnb{a}} = 
     \E{\ip{\Gdiff_{\tnb{a}}\En(\tnb{q};\uu,\tnb{a})}{\tnb{b}-\tnb{a}}}\geq 0
       \quad\forall \tnb{b}\in\rsP_n$,
\end{tabular}
\right.
\label{eq:rWeak}
\end{equation}
where we recall that $\Gdiff_{\uu}\En$ and $\ip{\cdot}{\cdot}$ refer to the {partial}
G\^{a}teaux derivative of $\En$ and the dual pairing between {the deterministic spaces}
$\sU^*$ and $\sU$, respectively, 
{whereas $\Gdiff_{\uu}\Ens$ is the partial G\^{a}teaux derivative of the new functional
$\Ens$, and $\ipd{\cdot}{\cdot}:=\E{\ip{\cdot}{\cdot}}$ is the duality pairing
between the stochastic spaces $\rsU^*$ and $\rsU$}. 
As an example pointing towards computation, from the 
first relation in \eqref{eq:rWeak} one can obtain
a ``strong version'' w.r.t\ $\rsU$, which applies $\P$-almost surely ($\P$-a.s.): 
\begin{equation}  \label{eq:rStrong-u}
   \ip{\Gdiff_{\uu}\En(\tnb{q}(\theta);\uu(\theta),\tnb{a}(\theta))}{\vv(\theta)}  = 0 
       \quad \forall \vv\in\rsU \quad \P-\text{a.s.}
\end{equation}
As $\P$-a.s.\ one has $\uu(\theta) \in \sU_n$ and $\vv(\theta) \in \sU$ --- this
means that except for $\theta$ in a set of vanishing probability --- both the
constraint \eqref{eq:Un} and the first of the governing equations in \eqref{Weak} is
satisfied like in the deterministic case.  A similar statement can be made about the
second relation in \eqref{eq:rWeak} satisfying $\P$-a.s.\ the second relation in \eqref{Weak}.
These $\P-\text{a.s.}$ statements are possible due to the $\P-\text{a.s.}$
enforcement  of the conditions in the definition of $\rsU_n$ in \eqref{eq:rUn} 
and $\rsP_n$ in \eqref{eq:rPn}.

To be more specific, this general development is now applied to the definition of 
the functional from \eqref{RegVAF}, for the sake of simplicity without the Neumann boundary term.   
The perturbed functional is
\begin{multline}  \label{eq:rRegVAF}
   \En(\tnb{q}(\theta);\uu(\theta),\tnb{a}(\theta)) = 
   \int_{\Dom} \tns{g}(\tnb{a}(\theta,\x)) \Psi(\tnb{q}(\theta);\Vep(\uu(\theta,\x)))\,\di\x \\
   + \frac{G_c}{c_{\tns{w}}} \int_{\Dom}
 \left(\frac{\tns{w}(\tnb{a}(\theta,\x)))}{\ell}+\ell\ns{\nabla\tnb{a}(\theta,\x)}^2\right)\,\di\x 
\end{multline}
where we have assumed to perturb the elastic strain energy density $\Psi$. Then with \eqref{eq:rRegVAF} one has from \eqref{eq:rWeak}:
\[
  \ipd{\Gdiff_{\uu}\Ens(\uu,\tnb{a})}{\vv} =
   \E{\int_{\Dom} \tns{g}(\tnb{a}(\cdot,\x)) \dd_{\Vep} 
       \Psi(\tnb{q}(\cdot);\Vep(\rV{u}(\cdot,\x))) :
      \Vep(\rV{v}(\cdot,\x))\, \di\x} = 0   \quad \forall \rV{v} \in \rsU .
\]
In particular, for the example of anti-plane shear, this Gâteaux derivative, corresponding to 
\eqref{eq:Weak_aps-u} in the deterministic case, becomes
\[
\ipd{\Gdiff_{\uu} \Ens(\uu,\tnb{a})}{\vv} = 
  \E{\int_{\Dom} (1-\tnb{a}(\cdot,\x))^2\mu(\tnb{q}(\cdot))\,\nabla \uu(\cdot,\x)\cdot
            \nabla \vv(\cdot,\x) \,\di\x } ,
\]
where we have assumed a perturbation in the shear modulus $\mu$. Obviously, the ``strong version'' resulting from \eqref{eq:rStrong-u} could also be written
out for these particular cases; $\P$-a.s.\ for $\theta\in\Theta$ it just 
looks like the corresponding deterministic case.

\subsection{Some remarks about Young measures}
Here we briefly offer some remarks about Young measures (YM), which are used as a mathematical tool to generalise and relax variational formulations \cite{Roubicek1997book,Pedregal1997}, as well as ``statistical solutions'' \cite{FoiasTemam04}, lately
described by multi-point generalisations of YM \cite{Fjordholm2016a, fjordholm2016computation}, and their relation to the here proposed concept of ``stochastic solutions''.

Originally, the concept of YM was introduced to describe oscillation effects and was extended to include concentration effects (DiPerna-Majda measures \cite{DiPerna1987}) of minimising sequences in variational formulations which lack minimisers. As an example, one can consider 
\begin{align*}
\min_{u\in \Hp^1_0((0,1))} \int_{0}^1 u^2+[(u')^2-1]^2
\end{align*}
which generates 
approximate minimisers close to zero with finer and finer slopes, i.e.\ an oscillating derivative $u'=\pm 1$; however, such functions are not contained in the space $\Hp^1_0$. This can be relaxed with YMs, which 
are parametrised probability measures $\x\mapsto \nu_{\x}$ defined on the domain $\Dom$, and in the simplest case
can describe the accumulation points of minimising sequences $\{u_k\}_{k=1}^\infty$ of functions 
$u_k:\sY\rightarrow\sR$ in $\Lp_\infty(\Dom)$. Each $u_k$ generates for $\x\in\Dom$ a.e.\ a linear continuous functional for any continuous function $\varphi\in\Ck_{00}(\sR)$
with compact support: $\varphi \mapsto \varphi(u_k(\x)) = \ip{\updelta_{u_k(\x)}}{\varphi} =: \ip{\nu_{\x}^{(u_k)}}{\varphi}$,
i.e.\ the probability measure $\nu_{\x}^{(u_k)}:=\updelta_{u_k(\x)}$. The YMs arise when the sequence $\{u_k\}$ is bounded in $\Lp_\infty$ (and hence the $\{\nu_{\x}^{(u_k)}\}=\{\updelta_{u_k(\x)}\}$ are ``tight''), as then there is a weak* convergent sub-sequence (for the sake of simplicity with unchanged indices)
$u_k\stackrel{*}{\rightharpoonup} u$, such that the sequence $\{\varphi(u_k)\}$ also converges weak*.  

The weak*-limit 
\[
 \varphi(u_k) = \left[\x \mapsto \ip{\nu_{\x}^{(u_k)}}{\varphi} \right] \,
 \stackrel{*}{\rightharpoonup} \, \left[\x \mapsto \int_{\sR} \varphi(\y)\, \nu_{\x}^{(u)}(\di\y) =: \dual{\nu_{\x}^{(u)}}{\varphi} =: 
 \varphi_u(\x) \right] \in \Lp_\infty(\sY)
\]
defines again a linear continuous functional and hence the YM $\nu_{\x}^{(u)}$ (the ``narrow'' limit of the 
$\{\nu_{\x}^{(u_k)}\}=\{\updelta_{u_k(\x)}\}$), a measurable system of parametrised probability measures.

As the point-wise accumulation points of $\{\varphi(u_k(\x))\}$ may be different from $\varphi(u(\x))$,
this may be overcome with the YM generated by the sequence, which describes the
limit as $\varphi_u(\x) = \dual{\nu_{\x}^{(u)}}{\varphi}$, which may be seen as the expected value of $\varphi$ w.r.t.\ the probability measure $\nu_{\x}^{(u)}$.

YM have limitations to describe fracture and damage fields.
The YM corresponding to a damage (or fracture) field, may describe the probability distribution $\nu_\x$ of the damage at a particular point $\x$. However, information about spatial correlations is lacking in such a setting. Particularly, probabilities that the crack appears in a point $\y$ conditional on a crack at another point $\x$ cannot be described.

These limitations can be overcome with RFs as ``stochastic solutions''.  If we have a RF $\tnb{a}\in \rsP = \C{S}\otimes\sP = \Lp_2(\Theta)\otimes\Hp^1(\Dom)
\cong \Lp_2(\Theta;\Hp^1(\Dom))$, similarly as before for any
$\varphi\in\Ck_{00}(\sR)$ one has that for
$\x\in\Dom$ a.e.\ the expression $\E{\varphi\circ\tnb{a}(\cdot,\x)}=
 \E{\varphi(\tnb{a}(\cdot,\x))}$ is a linear continuous functional
of $\varphi$, and thus defines a probability measure (a YM) $\nu_{\x}^{(\tnb{a})}$ with
the same properties as before. 

It is the
distribution of the RV $\tnb{a}(\cdot,\x)$,
and if the ``stochastic solution'' is deterministic, i.e.\ the RF $\theta \mapsto \tnb{a}(\theta,\x) = \alpha(\x)$
is constant, then $\nu_{\x}^{(\tnb{a})} = \updelta_{\alpha(\x)}$. 
But a RF $\tnb{a}$ can also describe all desired kinds of $n$-point correlation measures
$\dual{\nu_{\x_1,\dots,\x_n}^{(\tnb{a})}}{\varphi} = \E{\varphi\circ(\prod_{i=1}^n\tnb{a}(\cdot,\x_i))}$
for $\x_1,\dots,\x_n \in \Dom$, and is thus much more informative than a YM,
hence the ``stochastic solutions'' here have at least the same descriptive power as the ``statistical solutions'' in \cite{Fjordholm2016a,fjordholm2016computation}, which are charaterized by such families of correlation measures. 

\subsection{Numerical computations}
\label{sec:numerics}

Section \ref{sec:stochastic-formulations-of-variational-problems} has outlined a stochastic reformulation of variational problems. In this section, we will introduce sampling approximations for numerical computations and illustrate the procedure with the anti-plane shear test case. Before proceeding, we summarize our strategy to obtain the QoIs as follows:  
\begin{enumerate}
  \item perturbation of the parameters $\Vq$ of the problem by a zero-mean RV $\tilde{\tnb{q}}:\Theta\rightarrow\sQ$ (e.g. $\tilde{\tnb{q}}\in \sL_2(\Theta;\sQ)$) and a positive parameter $\eta$, i.e. the parameters of the problem are considered as
  \begin{align}
  \label{eq:q_perturb}
  \tnb{q}_\eta(\theta)=q_0+\eta \tilde{\tnb{q}}(\theta).
  \end{align}
  
  \item Use of a (numerical) solver to obtain a realization of the output RV $\vek{\Upsilon}_{\eta}(\uu(\theta),\tnb{a}(\theta))$.
  
  \item 
  Numerical computation of the QoIs $\E{\vek{\Upsilon}_\eta(\uu,\tnb{a})}$, e.g. mean value or variance, through a sampling method. 
 \end{enumerate}

Note that the variational problems may have multiple solutions, but a numerical solver provides a unique solution for each parameter. This allows us to make use of RVs in the usual way.
Thanks to the perturbation, the resulting algorithm still provides a RV. Yet, the influence of the solver on the numerical algorithm and results remains questionable. Note also that the solution operator may be discontinuous as a small change in the parameters may lead to a significantly different response. This makes the functional approximation of the solution operator complicated and an approximation, which allows to describe discontinuities, is required. In this sense, the Monte Carlo method is a sensible choice.

The theoretical results of the previous section will now be illustrated returning to the anti-plane shear test. Following the approach outlined so far we first introduce the random perturbations before applying the Monte Carlo algorithm. 

\subsubsection{Random perturbations}
In this case, among many possible stochastic inputs, we focus on a probabilistic modeling of the geometry. Note that a stochastic geometry is typically handled by transforming back the perturbed domain to a deterministic reference domain and the weak formulation is expressed on this reference domain using pull-back operators. After applying the pull-back, only the material tensors are random. Hence, a random geometry setting is closely related to Section~\ref{sec:stochastic-formulations-of-variational-problems}, where a perturbation of the material parameter on a fixed domain was considered for illustration purposes. In the context of the mechanical problem considered here, it is expected that the shape of the hole will have a significant influence on the formation of the crack path. 

Therefore, we perturb the hole geometry using a rather general approach for star-shaped objects, which is employed in inverse problems and uncertainty studies, see \cite{hiptmair2018large} for instance.
Particularly, the hole geometry $\{\x(r,\varphi)=(r\cos \varphi, r\sin \varphi)\,|\, \text{ for }r=R>0 \text{ and }\varphi\in[0,2\pi)\}$ expressed in polar coordinates is randomised by perturbing its radius
\begin{align}
    \tnb{q}_\eta(\theta) &= R + \eta \tilde{\tnb{q}}(\theta)
    &
    \text{with}\quad \tilde{\tnb{q}}(\theta)=\sum_{j=1}^J c_j \rv{y}_{2j-1}(\theta) \cos(j \varphi) + s_j \rv{y}_{2j}(\theta) \sin(j \varphi),
    \label{eq:random_perturbation}
\end{align}
where $R$ refers to the unperturbed (nominal) radius of a circle, $\eta$ controls the magnitude of the perturbation and $c_j,s_j$ are deterministic coefficients which are used to weight the influence of the different harmonics. In particular we set $c_j=s_j=\frac{1}{j}$ such that the influence of higher harmonics is successively decreasing. Finally, in \eqref{eq:random_perturbation}, $\rv{y}_{k}:\Theta \rightarrow \mathbb{R}, k=1,\ldots,2J$ are assumed to be uniformly distributed RVs on the interval $[-1,1]$. We additionally assume that the RVs are independent of each other. The parameterization introduced through $\rv{y}_i, i=1,\ldots,2J$ enables the generation of RF samples by simply drawing uniformly distributed pseudo-random numbers. It should be further noted that \eqref{eq:random_perturbation} represents a Karhunen-Lo\`eve expansion, introduced in Section~\ref{sec:stochastic-formulations-of-variational-problems}, which is more commonly known as proper orthogonal decomposition in the context of reduced order modeling. Note that we now use the Karhunen-Lo\`eve expansion for the input data.

\begin{figure}[t!]
\centering
\begin{tikzpicture}
\node at (-5,4.5) {\includegraphics[scale=0.2]{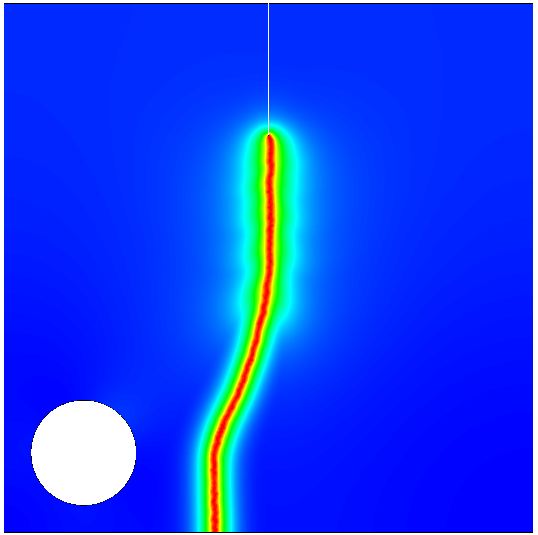}};
\node at (0,4.5) {\includegraphics[scale=0.2]{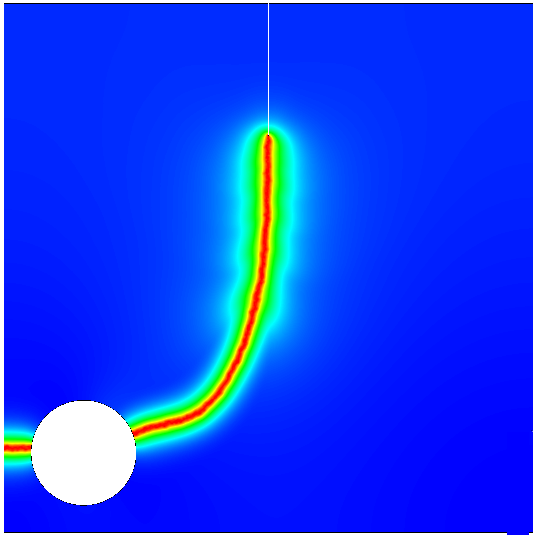}};
\node at (5,4.5) {\includegraphics[scale=0.2]{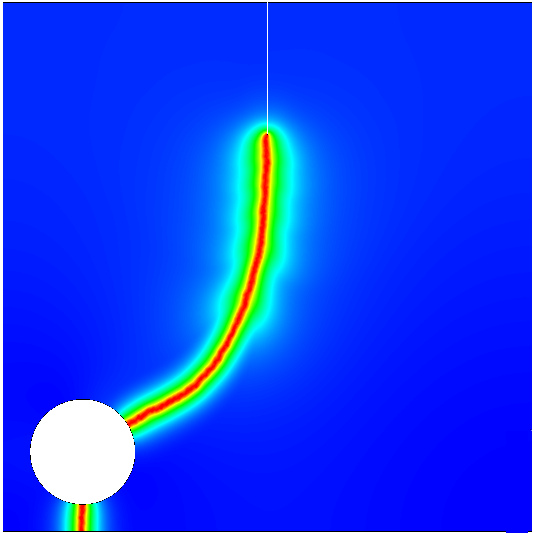}};
\node at (-5,0) {\includegraphics[scale=0.2]{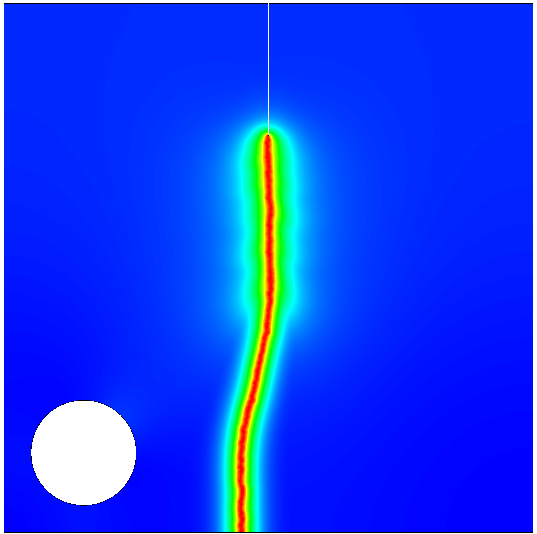}};
\node at (0,0) {\includegraphics[scale=0.2]{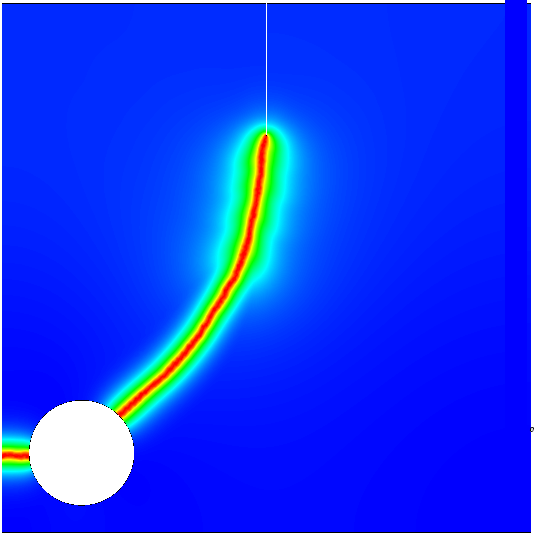}};
\node at (5,0) {\includegraphics[scale=0.2]{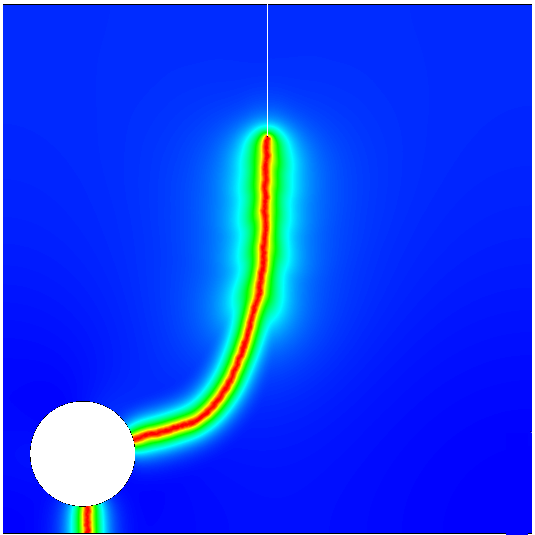}};
\node at (-5,-4.5) {\includegraphics[scale=0.2]{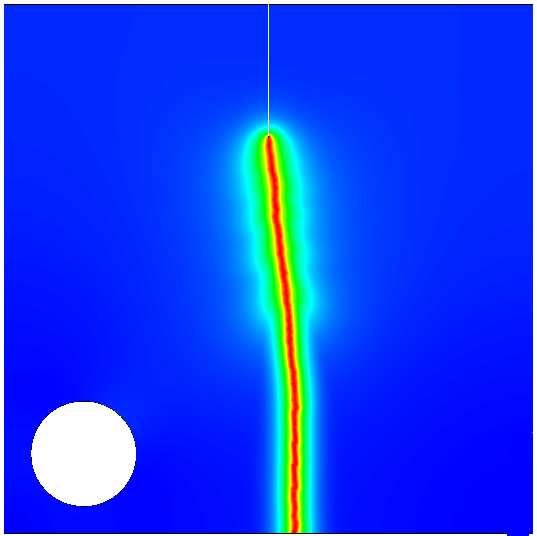}};
\node at (0,-4.5) {\includegraphics[scale=0.2]{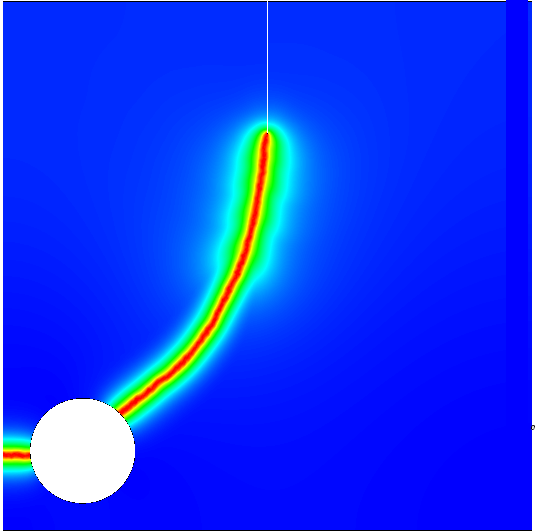}};
\node at (5,-4.5) {\includegraphics[scale=0.2]{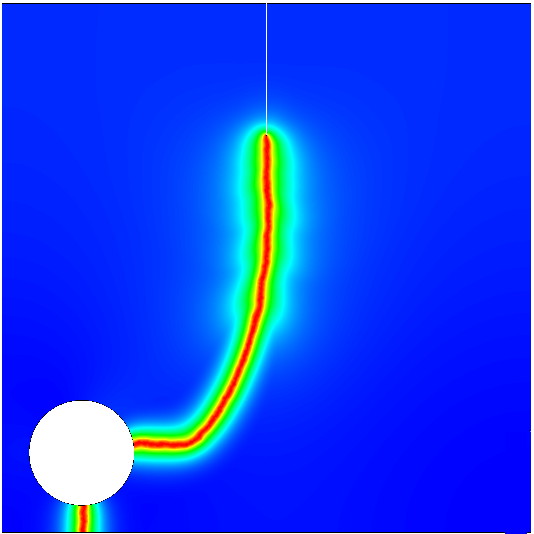}};
\end{tikzpicture}
\caption{Different realizations of the Monte Carlo simulation. All three characteristic crack patterns from the deterministic simulations appear, see Table \ref{tab:CrackClass}.}
\label{fig:single_real}
\end{figure}

In a specific application scenario, \eqref{eq:random_perturbation} can be used to model uncertainties in the geometry due to manufacturing imperfections.  Measurement data, based on imaging for instance, could then be used to infer the probability distribution of the coefficients $c_j$ and $s_j$ and possible correlations. These aspects are, however, not elaborated in any detail here.

\subsubsection{Computing statistical moments}
We employ the Monte Carlo method based on a sample $\{\tnb{q}_\eta^{(i)}\}_{i=1}^M$ of the perturbed parameter. Such a sample can be obtained with separated representations, the Karhunen-Lo\`eve expansion in particular, which has already been introduced in Section~\ref{sec:stochastic-formulations-of-variational-problems} and will be illustrated below through a specific example. We then employ a numerical solver to obtain $\vek{\Upsilon}_{\eta,h}(\uu^{(i)},\tnb{a}^{(i)})$ , where $h$ denotes a discretization parameter. This allows to approximate QoIs, e.g.
\begin{align*}
    %\E{\vek{\Upsilon}_{\ep,h}(\uu,\tnb{a})}&\approx\frac{1}{M} \sum_{i=1}^M \vek{\Upsilon}_{\ep,h}(\uu^{(i)},\tnb{a}^{(i)}), \\
 \mu := \E{\vek{\Upsilon}_{\eta,h}(\uu,\tnb{a})}&\approx \frac{1}{M} \sum_{i=1}^M \vek{\Upsilon}_{\eta,h}(\uu^{(i)},\tnb{a}^{(i)}) =: \mu_{\eta,h},
 \\
\MR{var}_{\vek{\Upsilon}_{\eta,h}(\uu,\tnb{a})}  &\approx \frac{1}{M} \sum_{i=1}^M (\vek{\Upsilon}_{\eta,h}(\uu^{(i)},\tnb{a}^{(i)}) - \mu_{\eta,h})^2.
\end{align*}

\begin{figure}[!t]
\begin{center}
\begin{tikzpicture}
\node at (0,0) {\includegraphics[width=0.7\textwidth]{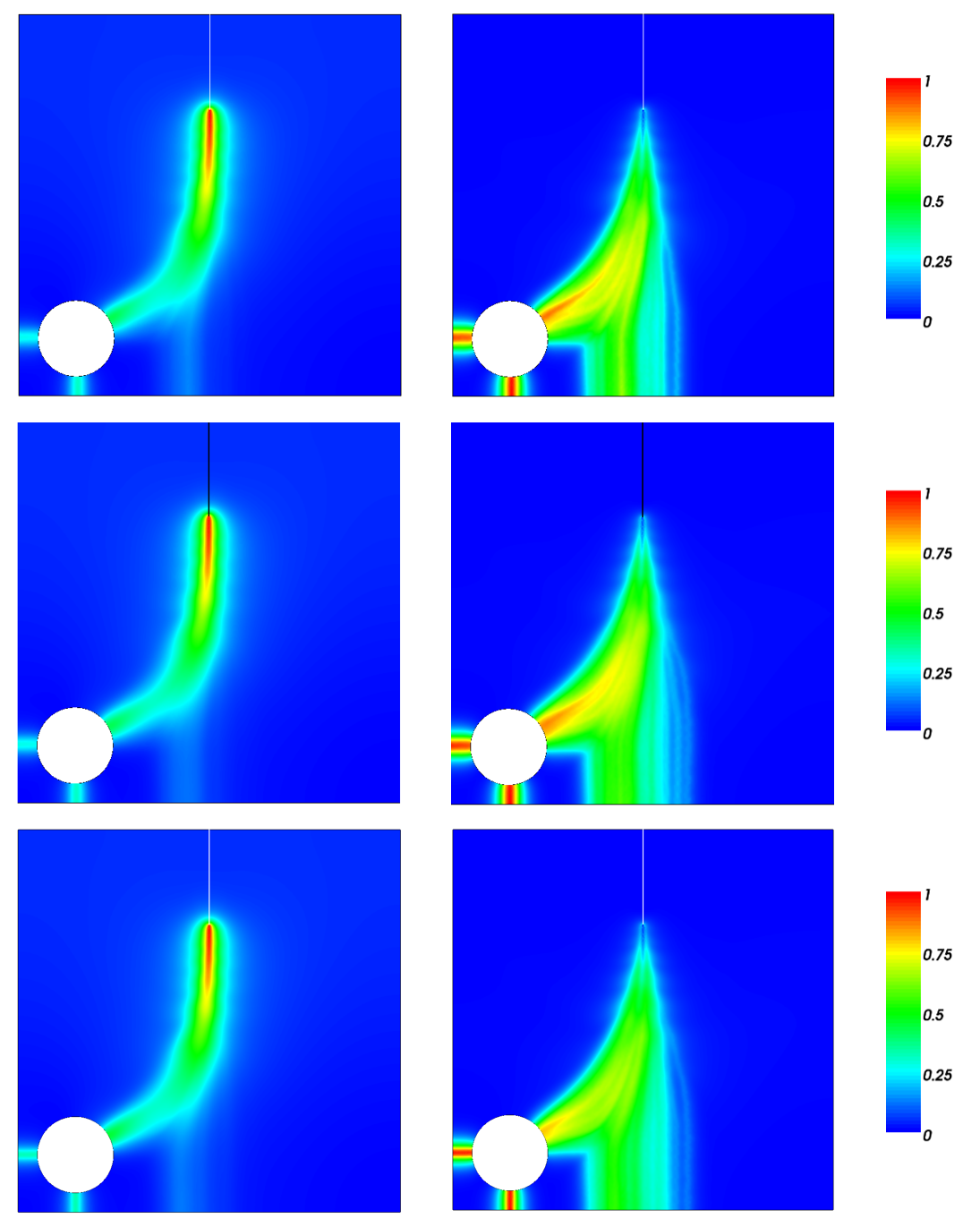}};
\node[fill=white,inner sep=1pt] at (-7.5,4.3) {$\eta=0.02, M=200$};
\node[fill=white,inner sep=1pt] at (-7.5,-0.4) {$\eta=0.02, M=400$};
\node[fill=white,inner sep=1pt] at (-7.5,-4.9) {$\eta=0.01, M=400$};
\node at (-3,7.5) {$\mu_{\tnb{a}}$};
\node at (1.8,7.5) {$\text{std}_{\tnb{a}}= \sqrt{\MR{var}_{\tnb{a}}}$};
\end{tikzpicture}
\end{center}
\caption{Expected value $\mu_{\tnb{a}}$ and standard deviation $\text{std}_{\tnb{a}}= \sqrt{\MR{var}_{\tnb{a}}}$ of the phase field for different magnitudes of $\eta$ and $M$.}
\label{fig:eps_moments}
\end{figure}
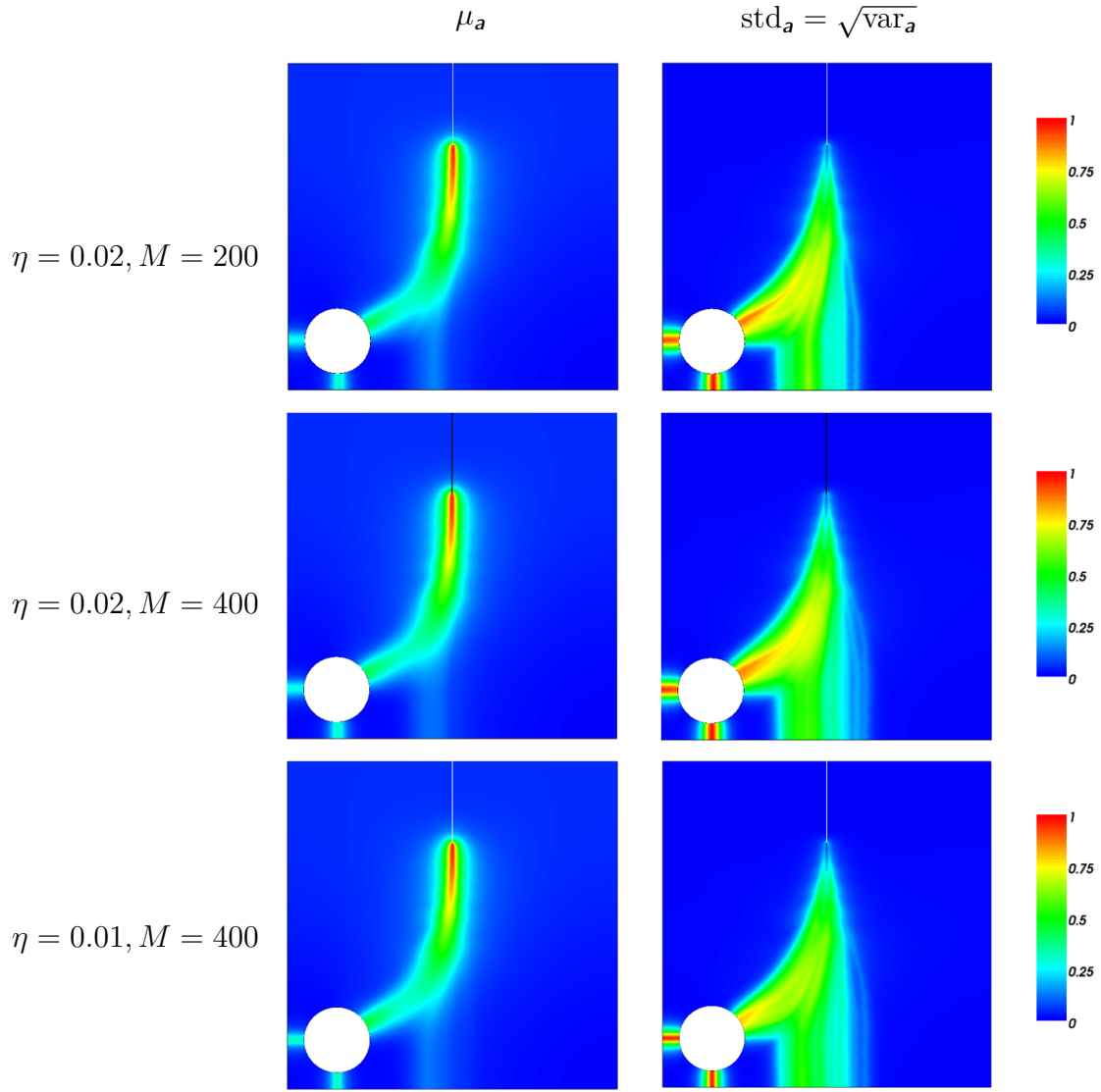

We now report the stochastic numerical results for our anti-plane shear test. This subsection focuses on the mean value and variance of the phase field, whereas computing crack pattern probabilities is treated in the next subsection. We employ Mesh 2 with $(h_\mathtt{min},h_\mathtt{max})=(\frac{1}{4}\ell,\ell)$ and the radius of the hole given by (\ref{eq:random_perturbation}). The applied displacement is given by $\bar{\Vu}_n=n\Delta\bar{u}$, $n=1,...,15$ with $\Delta\bar{\Vu}=0.1$. Results shown (final crack pattern) will be those of the last loading step.% $\bar{u}_{15}=1.5$.

Figure \ref{fig:single_real} depicts the crack phase field solution $\alpha^{(i)}$ at nine different realizations $i_1, i_2,$ \dots, $i_9 \in \{1,\ldots,M\}$. These realizations illustrate the strong variability of the crack pattern which is triggered by randomly perturbing the geometry. 
%In the plots the distortion of the hole geometry, which has been obtained with $\eta = 0.01$ and $J=5$, is also visible. 
Note that all three crack patterns obtained in the deterministic simulations of Section \ref{SecPF} by perturbing the finite element mesh and classified in Table \ref{tab:CrackClass} are re-obtained here. However, crack types 2 and 3 show little deviations in different realizations, wheres crack type 1 covers a wider range of geometries, the final portion of the crack being located within a band close to the center of the specimen.

In Figure~\ref{fig:eps_moments} (top) we plot the estimated expected value and the standard deviation of the phase field for a perturbation amplitude of $\eta=0.02$ and a sample size $M=200$. Clearly, both the expected value and the standard deviation feature all three crack patterns observed already in Figure~\ref{fig:single_real}. The expected value of the phase field is particularly high directly at the notch where each crack begins. Among the three observed paths, the lowest values are obtained along the crack pattern remote from the hole (Type 1), which can be explained by the stronger spatial scattering in this area causing a smearing out of the phase field mean value. Conversely, the larger amplitudes for Type 2,3 cracks are attributed to the localization of the paths along the holes. Also the standard deviation is observed to be more concentrated for crack patterns passing through the hole (Type 2 and 3), whereas the band appearing in the center of the specimen for the Type 1 crack confirms the larger variability of this crack path.
Figure~\ref{fig:eps_moments} (middle) contains the same quantities with an increased sample size of $M=400$. Visually there is not much difference between the two plots. Increasing the sample size seems to lead to a more homogeneous mean and variance crack field along the vertical line in the middle of the domain.

For the largest sample size, i.e.\ $M=400$, we report the same numerical results, however, for a different perturbation magnitude $\eta=0.01$ in Figure~\ref{fig:eps_moments} (bottom). The crack distributions largely resemble those obtained with $\eta=0.02$. Differences can be identified in particular for the variance in the vicinity of the hole, yet this may represent a statistical effect, in view of the given sample size. Results for even smaller perturbation sizes (up to a minimum tested value of $\eta= 0.0025$) are not reported, since they show a qualitatively similar behavior.

\subsubsection{Computing crack pattern probabilities}

\begin{figure}[t!]
\begin{minipage}{0.49\textwidth}
\vspace*{3.7em}
 \subfloat[]{
\begin{tikzpicture}
\node at (-3.5,-1) {\color{white} aa};
\node at (0,0) {\includegraphics[scale=.32]{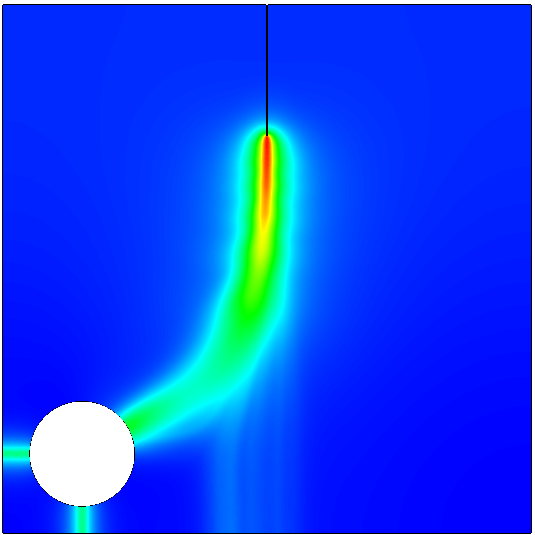}};
\draw[-,thick] (-2.24,0) -- (0.75,-2.25);
\node at (-2.85,0) {$s=0$};
\node at (0.75,-2.5) {$s=1$};
\draw[->,thick] (-2.24,0) -- (-1.6,-0.5);
\node at (-1.8,-0.15) {$s$};
\end{tikzpicture}
}
\end{minipage}
 \subfloat[]{
\begin{minipage}{0.485\textwidth}
\begin{tikzpicture}[scale = 0.85]
\begin{axis}[legend pos = north west, xlabel = line coordinate $s$, ylabel= Max. phase field probability density]
\addplot[color = black,thick,dashed] table [x=x, y=y1, col sep=comma] {images/densities_along_line_eps_0.0025_I200.csv};
\addplot[color = blue,thick,dashed] table [x=x, y=y2, col sep=comma] {images/densities_along_line_eps_0.0025_I200.csv};
\addplot[color = red,thick,dashed] table [x=x, y=y3, col sep=comma] {images/densities_along_line_eps_0.0025_I200.csv};
\addplot[color = green,thick] table [x=x, y=yall, col sep=comma] {images/densities_along_line_eps_0.0025_I200.csv};
\legend{crack 1, crack 2, crack 3, all cracks}
\end{axis}
\end{tikzpicture}
\end{minipage}
}
\caption{(a) Mean value of the phase field variable for $\eta=0.0025$ and $M=200$. Indicated is also the line $y = -\frac{2}{3} x + 1$ within the domain $D=[0,2] \times [0,2]$ with coordinate $s$, and (b) Probability density functions of the maximum phase field value along coordinate $s$ for all crack types.}
    \label{fig:densities_along_line}
\end{figure}
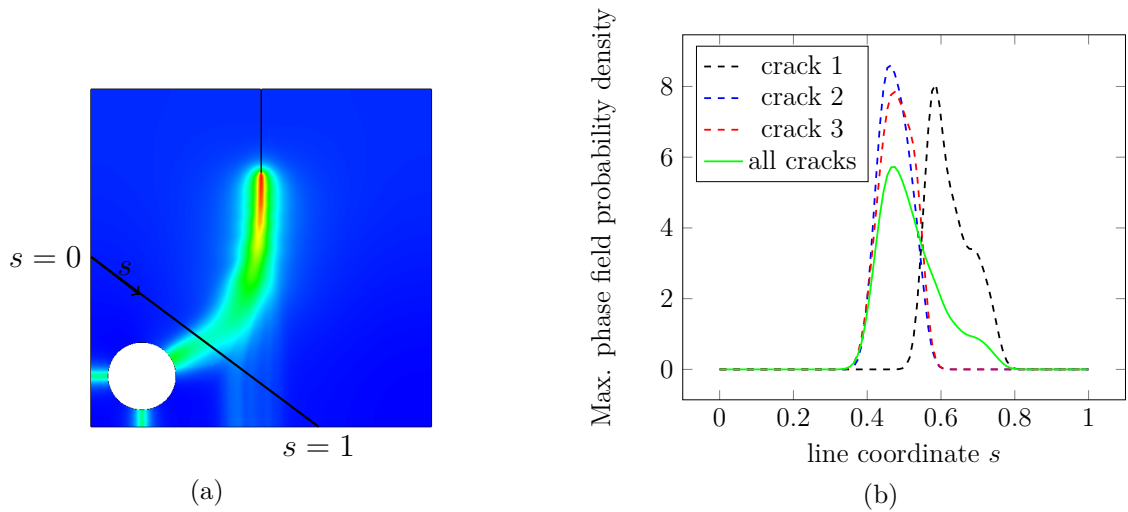

We recall that three different crack types occur for this application, referred to as crack $1,2,3$ see Table~\ref{tab:CrackClass}. 
To compute the corresponding probabilities of occurrence, the range of the phase field random variable $\tnb{a}(\theta)$ is clustered into three disjoint domains, i.e.\ $ \tnb{a}(\Theta)\in\bigcup_{i=1}^3 C_i\subset \sP_n$, corresponding to the three cracks, where $n$ represents the final loading step. Then, the probabilities $\P(\tnb{a}\in C_i)=\E{\ch{C_i}\circ \tnb{a}}$ for $i=1,2,3$ are numerically approximated with a sample as $p_i=\frac{1}{M}\sum_{m=1}^M \ch{C_i}\circ \tnb{a}^{(m)}$.
\begin{comment}
\begin{table}[h!]
    \centering
    \begin{tabular}{c|c|c|c|c}
         $\epsilon$ & $I$ & $P_1$ & $P_2$ & $P_3$ \\
         \hline
         0.0025 & 100 & 0.3 & 0.38 & 0.32 \\
         \hline
         0.0025 & 200 & 0.2750 & 0.365 & 0.36 \\
         \hline
         0.0025 & 400 & 0.26 & 0.375 & 0.365 \\
         \hline
         \hline
         0.005 & 100 & 0.38 & 0.33 & 0.29 \\
         \hline
         0.005 & 200 & 0.335 & 0.32 & 0.345 \\
         \hline
         0.005 & 400 & 0.315 & 0.315 & 0.37 \\
         \hline
         \hline
         0.01 & 100 & 0.36 & 0.33 & 0.31 \\
         \hline
         0.01 & 200 & 0.325 & 0.335 & 0.34 \\
         \hline
         0.01 & 400 & 0.31 & 0.34 & 0.35 \\
         \hline
         \hline
         0.02 & 100 & 0.27 & 0.36 & 0.37 \\
         \hline
         0.02 & 200 & 0.3 & 0.33 & 0.37 \\
         \hline
         0.02 & 400 & 0.3 & 0.33 & 0.37 
    \end{tabular}
    \caption{Monte Carlo probabilities of crack propagation for different crack paths.}
    \label{tab:probabilities}
\end{table}
\end{comment}
For a perturbation magnitude of $\eta=0.01$ and $M=200$, we obtain $p_1=0.325$, $p_2=0.335$, and $p_3=0.340$. Hence, the probabilities seem to be comparable, of approximately $1/3$, for each crack type. These numbers do not change significantly when the perturbation is varied. For instance, if $\eta=0.02$ and $M=200$, we obtain $p_{1} = 0.3$, $p_{2} = 0.33$, and $p_{3} = 0.37$. 

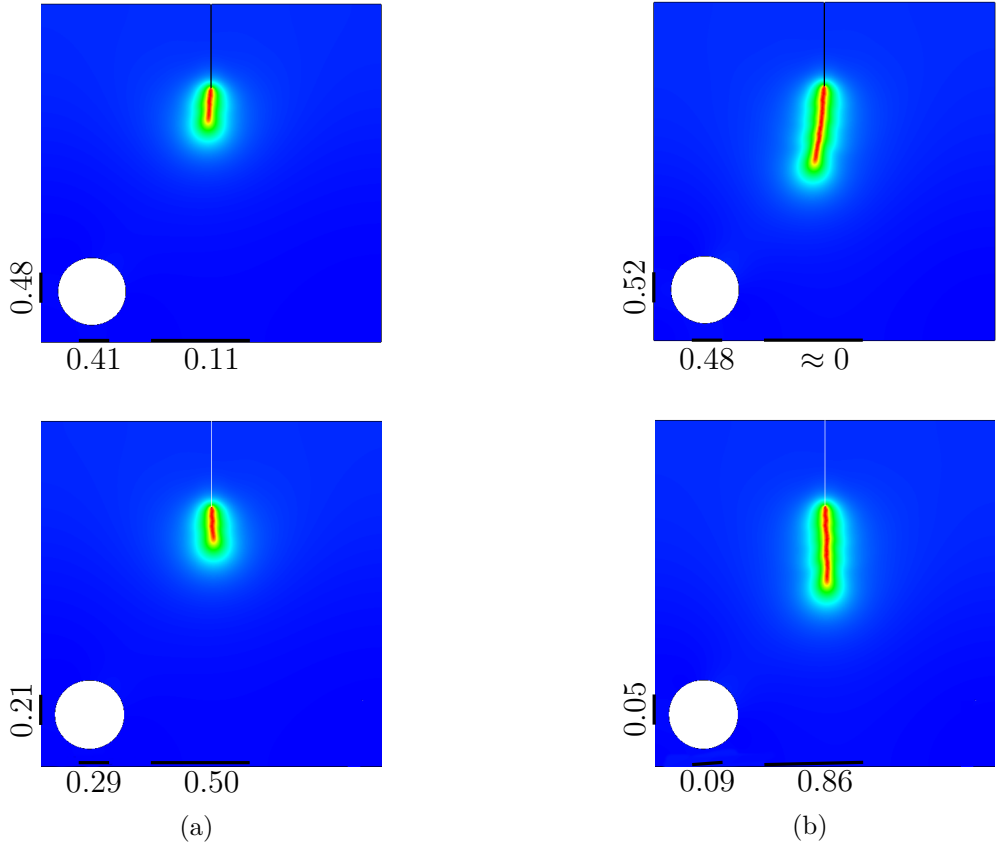
\begin{figure}[t!]
\begin{minipage}{0.49\textwidth}
\centering
\begin{tikzpicture}
\node at (-3.1,0) {\color{white} aaa};
\node at (0,0) {\includegraphics[scale=0.32]{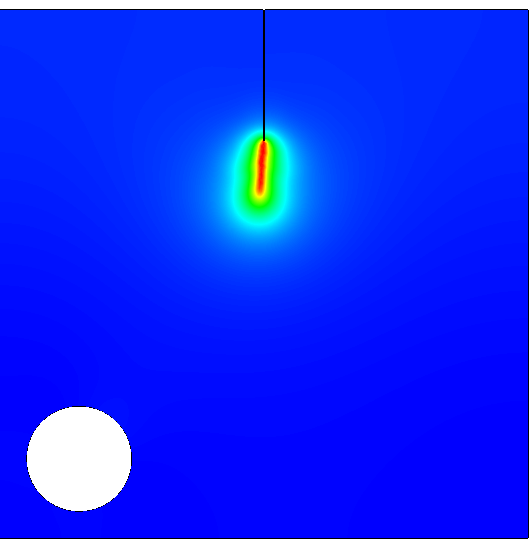}};
\draw[-,very thick] (-0.8,-2.25) -- (0.5,-2.25);
\draw[-,very thick] (-1.75,-2.25) -- (-1.35,-2.25);
\draw[-,very thick] (-2.25,-1.75) -- (-2.25,-1.35);
\node at (0,-2.5) {$0.11$};
\node at (-1.55,-2.5) {$0.41$};
\node[rotate=90] at (-2.5,-1.55) {$0.48$};
\end{tikzpicture}
~\newline
\subfloat[]{
\begin{tikzpicture}
\node at (0,0) {\includegraphics[scale=0.245]{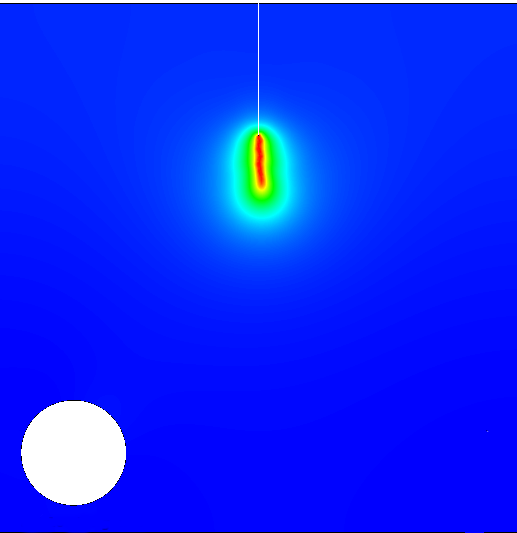}};
\draw[-,very thick] (-0.8,-2.25) -- (0.5,-2.25);
\draw[-,very thick] (-1.75,-2.25) -- (-1.35,-2.25);
\draw[-,very thick] (-2.25,-1.75) -- (-2.25,-1.35);
\node at (0,-2.5) {$0.50$};
\node at (-1.55,-2.5) {$0.29$};
\node[rotate=90] at (-2.5,-1.55) {$0.21$};
\end{tikzpicture}
}
\end{minipage}
\begin{minipage}{0.49\textwidth}
\centering
\begin{tikzpicture}
\node at (-3.1,0) {\color{white} aaa};
\node at (0,0) {\includegraphics[scale=0.32]{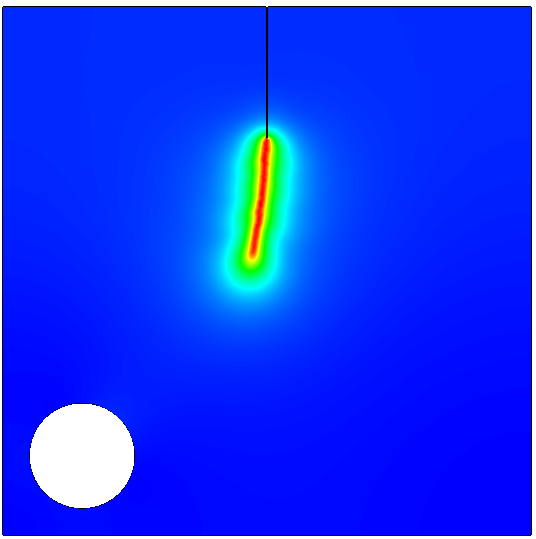}};
\draw[-,very thick] (-0.8,-2.25) -- (0.5,-2.25);
\draw[-,very thick] (-1.75,-2.25) -- (-1.35,-2.25);
\draw[-,very thick] (-2.25,-1.75) -- (-2.25,-1.35);
\node at (0,-2.5) {$\approx 0$};
\node at (-1.55,-2.5) {$0.48$};
\node[rotate=90] at (-2.5,-1.55) {$0.52$};
\end{tikzpicture}
~\newline
\subfloat[]{
\begin{tikzpicture}
\node at (0,0) {\includegraphics[scale=0.245]{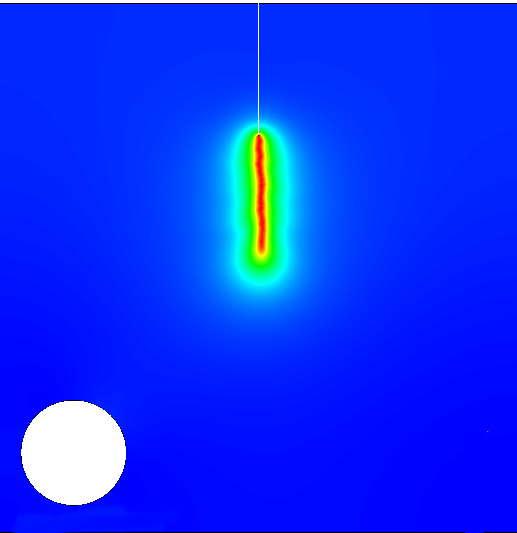}};
\draw[-,very thick] (-0.8,-2.28) -- (0.5,-2.25);
\draw[-,very thick] (-1.75,-2.28) -- (-1.35,-2.25);
\draw[-,very thick] (-2.25,-1.75) -- (-2.25,-1.35);
\node at (0,-2.5) {$0.86$};
\node at (-1.55,-2.5) {$0.09$};
\node[rotate=90] at (-2.5,-1.55) {$0.05$};
\end{tikzpicture}
}
\end{minipage}
\caption{Particular realizations of phase field variables at two different loading steps for $\eta=0.0025$ and $M=200$. The numbers refer to computed conditional crack probabilities for each different crack type, based on a Monte Carlo sample. (a) Cracks at loading step $n=9$ and (b) cracks at loading step $n=11$.}
\label{fig:conditional_probabilities}
\end{figure}

We proceed by discussing the possibility of updating probabilities during crack propagation. This could be useful to analyze and assess initiated but not fully developed cracks. To this end, conditioning on a given damage state is required. We simplify the computations by considering a cut of the domain, see Figure~\ref{fig:densities_along_line} on the left. 

We assume that all cracks pass through a line \ $L=\{\x\in D \mid \x = \vek{p} + s \vek{u}, s \in [0,1] \}$, which is represented by a RV $\rv{y}: \Theta\rightarrow L$ defined as
$\rv{y}(\theta)=\argmax_{\x\in L} \{\tnb{a}_n(\theta,\x) \mid, n \geq 1 \}$.
 The probability of a particular intersection point of the crack with the line is characterised with the probability density $f_\rv{y}:L\rightarrow[0,\infty).$ Here, we approximate the density with kernel estimation based on the sample $\{\rv{y}^{(i)}\}_{i=1}^M$, see Figure~\ref{fig:densities_along_line}. Moreover, based on the partitioning of the range of $\tnb{a}$, we can also estimate the conditional probability densities $f_{\rv{y}| C_i}$ for individual cracks. We observe that the conditional densities for cracks $2,3$ nearly coincide, which is expected, since the domain is cut before the cracks reach the hole where it is almost impossible to distinguish between both crack patterns.

Next, we address the computation of conditional probabilities, i.e.\ the probability that crack $C_1$ happens when the intersection of the crack with a line $L$ is observed at a position $x_c$.
Mathematically, this computation builds on Bayes' theorem
\begin{align}
\P(\tnb{a}\in C_1 | \rv{y}=x_c) = \frac{f_{\rv{y}|C_1}(x_c)\P(\tnb{a}\in C_1)}{f_\rv{y}(x_c)},
\end{align}
where $\P(\tnb{a}\in C_1)=p_1$ is the total probability that crack $C_1$ occurs, $f_\rv{y}(x_c)$ is the probability of the crack at point $x_c$ (obtained from the green line in Figure~\ref{fig:densities_along_line}), and $f_{\rv{y}\mid C_1}(x_c)$ is the conditional probability of a crack intersecting at $x_c$ for a crack of type $C_1$ (obtained from the dashed black line in Figure~\ref{fig:densities_along_line}). Figure~\ref{fig:conditional_probabilities} shows cracks at different stages of their evolution, together with the associated crack type probabilities. The upper plots depict a crack through the hole for $n=9$ and $n=11$, respectively. Already after bending slightly to the left (upper left plot) the probability of a obtaining a type $1$ crack, not passing through the hole, is significantly reduced. After two additional loading steps (upper right plot) the associated probability is almost zero. The two plots on the bottom represent a type $1$ crack, again at loading stages $n=9$ and $n=11$, respectively. This crack initially is slightly deviated to the right, which causes the type 1 probability to increase. At loading step $n=11$ (bottom right plot) the probability of passing through the hole is already very low. These examples show the large influence of conditioning the crack probabilities to partially developed crack patterns. We emphasize that these computations critically rely on correlation information for the phase field at different points in the computational domain. Such information is naturally available in the stochastic solution setting of the paper.

\section{Conclusions}
\label{Conclusions}
The phase field approach can cope with several challenges in the computational treatment of brittle fracture. However, its solution is non-unique in general, and this raises the question of the meaning and representativeness of the (possibly several) solutions which can be found numerically. Through an illustrative test case we have exposed the non-uniqueness of the solution due to competing energy levels of a variety 
of crack paths. In particular, the crack propagation was found to be highly sensitive to meshing and geometric perturbations. No other types of perturbations were tested.

In view of these findings, we have introduced the concept of a ``stochastic solution'' in the form of random fields as a new and 
general concept in brittle fracture, allowing one to obtain various crack patterns and their probabilities. 

Random fields can capture correlations, which is not possible with standard Young measures for instance. 
They further permit to condition on not fully developed crack paths, which is interesting from a practical
perspective. The stochastic solution concept does not remove the problem of non-uniqueness in general: one 
may obtain all possible crack patterns at once, but 
with possibly non-unique probabilities (see e.g. the example in Section \ref{sec:introductory-concepts}). However, the computed probabilities  in some numerical experiments were comparable, despite the adoption of different solution approaches (see the global minimization for the sharp crack model in Section \ref{sec:1D_griffith} and the local minimization with the regularized model in Section \ref{secPF_1D}). Moreover, further numerical tests (not reported in the paper) seem to indicate that, when inducing the geometrical perturbation adopted in Section \ref{sec:numerics}, the sensitivity of the stochastic solution, expressed in terms of probability distribution of the different crack paths, with respect to changes in the numerical setup of the problem is lower than the sensitivity of the deterministic solution to such changes. Importantly, the numerical
results further showed a dependence of the computed crack probabilities on the type of perturbation (see the three examples in Section \ref{sec:1D_griffith}), which is to be expected and also 
desired from a physical perspective. In general we conjecture the existence of a unique stochastic solution for a given 
physical perturbation and a formulation providing this unique solution is still to be developed.

%%%%%%%%%%%%%%%%%%%%%%%%%%%%%%%%%%%%%%%%%%%%%%%%%%%%%%%%%%%%%%%%%%%%%%%%%%%%%%%%%%%%%%%%%%%%%%%

\end{document}